\documentclass{article}

\usepackage[final]{neurips_2018}

\usepackage[utf8]{inputenc}
\usepackage[T1]{fontenc}
\usepackage{hyperref}       
\usepackage{booktabs}
\usepackage{amsfonts}
\usepackage{amsmath}
\usepackage{nicefrac}
\usepackage{microtype}
\usepackage{color, soul}
\usepackage{tikz,pgf}
\usetikzlibrary{positioning}
\usepackage{float}
\usepackage{subfig}
\usepackage{xr}
\usepackage{multirow}
\usepackage{array}
\usepackage{standalone}

\newenvironment{thma}[1]{\par\noindent{\bf Theorem #1\ }\em}{\em}

\newenvironment{prf}{\noindent\textit{Proof:}\begin{mdseries}}{\end{mdseries}{\hfill\scriptsize$\Box$}}

\newcommand{\G}{{\mathcal G}}
\newcommand{\E}{\mathbb{E}}

\newtheorem{thm}{Theorem}

\newtheorem{dfn}{Definition}

\DeclareMathOperator{\dis}{dis}

\DeclareMathOperator{\pa}{pa}
\DeclareMathOperator{\de}{de}

\DeclareMathOperator{\ch}{ch}
\DeclareMathOperator{\an}{an}

\DeclareMathOperator{\ant}{ant}

\DeclareMathOperator{\nd}{nd}

\DeclareMathOperator{\nb}{nb}

\DeclareMathOperator{\sib}{sib}

\title{Identification and Estimation Of Causal Effects from Dependent Data}

\author{
  Eli Sherman\\
  Department of Computer Science\\
  Johns Hopkins University\\
  Baltimore, MD 21218 \\
  \texttt{esherman@jhu.edu} \\
   \And
   Ilya Shpitser \\
   Department of Computer Science \\
   Johns Hopkins University \\
   Baltimore, MD 21218 \\
   \texttt{ilyas@cs.jhu.edu} \\
}

\begin{document}

\maketitle

\begin{abstract}
The assumption that data samples are independent and identically distributed (iid) is standard in many areas of statistics and machine learning. Nevertheless, in some settings, such as social networks, infectious disease modeling, and reasoning with spatial and temporal data, this assumption is false. An extensive literature exists on  making causal inferences under the iid assumption \cite{robins86new, pearl95causal, tian02onid, shpitser06id}, even when unobserved confounding bias may be present. But, as pointed out in \cite{shalizi2011homophily}, causal inference in non-iid contexts is challenging due to the presence of both unobserved confounding \emph{and} data dependence. In this paper we develop a general theory describing when causal inferences are possible in such scenarios. We use \textit{segregated graphs} \cite{shpitser15segregated}, a generalization of latent projection mixed graphs \cite{verma90equiv}, to represent causal models of this type and provide a complete algorithm for non-parametric identification in these models.  We then demonstrate how statistical inference may be performed on causal parameters identified by this algorithm. In particular, we consider cases where only a single sample is available for parts of the model due to \emph{full interference}, i.e., all units are pathwise dependent and neighbors' treatments affect each others' outcomes \cite{tchetgen2017auto}. We apply these techniques to a synthetic data set which considers users sharing fake news articles given the structure of their social network, user activity levels, and baseline demographics and socioeconomic covariates.
\end{abstract}

\section{Introduction}
\label{sec:intro}

The assumption of independent and identically distributed (iid) samples is ubiquitous in data analysis. In many research areas, however, this assumption simply does not hold. For instance, social media data often exhibits dependence due to homophily and contagion \cite{shalizi2011homophily}.
Similarly, in epidemiology, data exhibiting herd immunity 
is likely dependent across units.  Likewise, signal processing and sequence learning often consider data that are spatially \cite{mnih2015human} or temporally \cite{sutskever2014sequence} dependent.

In causal inference, dependence in data often manifests as \emph{interference} wherein some units' treatments may causally affect other units' outcomes \cite{hudgens08toward, ogburn14interference}.  Herd immunity is a canonical example of interference since other subjects' vaccination status causally affects the likelihood of a particular subject contracting a disease. Even under the iid assumption, making causal inferences from observed data is difficult  due to the presence of unobserved confounding. This difficulty is worsened when interference is present, as described in detail in \cite{shalizi2011homophily}.  In general, these difficulties prevent identification of causal parameters of interest, making estimation of these parameters from data an ill-posed problem.
An extensive literature on identification of causal parameters (under the iid assumption) has been developed.  The \emph{g-formula} \cite{robins86new} identifies any interventional distribution in directed acylcic graph-based (DAG) causal models without latent variables.  Pearl showed that in certain cases identification is possible even in the presence of unobserved confounding via the \emph{front-door criterion} \cite{pearl95causal}.
These results were generalized into a complete identification theory in hidden variable causal DAG models via the ID algorithm \cite{tian02onid,shpitser06id}.  An extensive theory of estimation of identified causal parameters has been developed.  Some approaches are described in \cite{robins86new,robins99marginal}, although this is far from an exhaustive list.  While work on identification and estimation of causal parameters under interference exists \cite{hudgens08toward,tchetgen12on,ogburn14interference, pena2018reasoning, pena2016learning, maier2013reasoning, arbour2016inferring}, no general  theory has been developed up to now.  In this paper, we aim to provide this theory for a general class of causal models that permit interference.

\section{A Motivating Example}
\label{sec:example}

To motivate subsequent developments, we introduce the following example application. Consider a large group of internet users, belonging to a set of online communities, perhaps based on shared hobbies or political views. For each user $i$, their time spent online
$A_i$ is influenced by their observed vector of baseline factors $C_i$, and unobserved factors $U_i$. In addition, each user maintains a set of friendship ties with other users via an online social network.  The user's activity level in the network, $M_i$, is potentially dependent on the user's friends' activities, meaning that for users $j$ and $k$, $M_j$ and $M_k$ are potentially dependent.  The dependence between $M$ variables is modeled as a stable symmetric relationship that has reached an equilibrium state. Furthermore, activity level $M_i$ for user $i$ is influenced by observed factors $C_i$, time spent online $A_i$, and the time spent online $A_j$ of any unit $j$ who is a friend of $i$.
Finally, we denote user $i$'s sharing behavior by $Y_i$.  This behavior is influenced by the social network activity of the unit,
and possibly the unit friends' time spent online.



A crucial assumption in our example is that for each user $i$, purchasing behavior $Y_i$ is causally influenced by baseline characteristics $C_i$, social network activity $M_i$, and unobserved characteristics $U_i$, but time spent online $A_i$ does not \emph{directly} influence sharing $Y_i$, except as mediated by social network activity of the users.
While this might seem like a rather strong assumption, it is more reasonable than standard ``front-door'' assumptions \cite{pearl09causality} in the literature, since we allow the entire social network structure to mediate the influence $A_i$ on $Y_i$ for every user.


We are interested in predicting how a counterfactual change in a set of users' time spent online influences their purchasing behavior.  Note that solving this problem from observed data on users as we described is made challenging both by the fact that unobserved variables causally affect both community membership and sharing, creating spurious correlations, and because social network membership introduces dependence among users.  In particular, for realistic social networks, every user's activity potentially depends on every other user's activity (even if indirectly).  This implies that a part of the data for this problem may effectively consist of a single dependent sample \cite{tchetgen2017auto}.


In the remainder of the paper we formally describe how causal inference may be performed in examples like above, where both unobserved confounding and data dependence are present.
In section \ref{sec:review} we review relevant terminology and notation, give factorizations defining graphical models, describe causal inference in models without hidden variables, and give identification theory for such models in terms of a modified factorization.  We also introduce the dependent data setting we will consider.  In section \ref{sec:hidden} we describe more general \emph{nested} factorizations \cite{richardson17nested} applicable to marginals obtained from hidden variable DAG models, and describe identification theory in causal models with hidden variables in terms of a modified nested factorization.  In section \ref{sec:cg}, we introduce causal chain graph models \cite{lauritzen02chain} as a way of modeling causal problems with interference and data dependence, and pose the identification problem for interventional distributions in such models.
In section \ref{sec:sg-id} we give a sound and complete identification algorithm for interventional distributions in a large class of causal chain graph models with hidden variables, which includes the above example, but also many others.  We describe our experiments, which illustrate how identified functionals given by our algorithm may be estimated in practice, even in \emph{full interference} settings where all units are mutually dependent, in section \ref{sec:exp}.  Our concluding remarks are found in section \ref{sec:conclusion}.

\begin{figure}[t]
	\begin{center}
		\begin{tikzpicture}[>=stealth, node distance=1.0cm]
		\tikzstyle{format} = [draw, very thick, circle, minimum size=5mm,
		inner sep=0pt]
		\tikzstyle{square} = [draw, very thick, rectangle, minimum size=3.8mm]
		
		\begin{scope}[xshift=0cm]
		\path[->, very thick]
		node[format] (a1) {$A_1$}
		node[format, above of=a1] (c1) {$C_1$}
		node[format, below of=a1] (m1) {$M_1$}
		node[format, below of=m1] (y1) {$Y_1$}
		
		node[format, right of=a1] (a2) {$A_2$}
		node[format, above of=a2] (c2) {$C_2$}
		node[format, below of=a2] (m2) {$M_2$}
		node[format, below of=m2] (y2) {$Y_2$}
		
		node[format, gray, xshift=0.4cm, left of=m1] (u1) {$U_1$}
		node[format, gray, xshift=-0.4cm, right of=m2] (u2) {$U_2$}
		
		(c1) edge[blue] (a1)
		(c1) edge[blue, bend right=40] (m1)
		(c1) edge[blue, bend right=65] (y1)
		(a1) edge[blue] (m1)
		(m1) edge[blue] (y1)
		
		(c2) edge[blue] (a2)
		(c2) edge[blue, bend left=40] (m2)
		(c2) edge[blue, bend left=65] (y2)
		(a2) edge[blue] (m2)
		(m2) edge[blue] (y2)
		
		
		
		
		(u1) edge[red] (a1)
		(u1) edge[red] (y1)
		(u2) edge[red] (a2)
		(u2) edge[red] (y2)
		
		(a1) edge[blue] (m2)
		(a2) edge[blue] (m1)
		
		(a1) edge[blue] (y2)
		(a2) edge[blue] (y1)
		
		(m1) edge[-, brown] (m2)
		node[below of=y1, yshift=0.2cm, xshift=0.5cm] (l) {$(a)$}
		;
		\end{scope}
		
		\begin{scope}[xshift=2.5cm]
		\path[->, very thick]
		node[format, gray] (c1) {$U_1$}
		node[format, below of=c1] (a1) {$A_1$}
		node[format, below of=a1] (y1) {$Y_1$}
		node[format, gray, right of=c1] (c2) {$U_2$}
		node[format, below of=c2] (a2) {$A_2$}
		node[format, below of=a2] (y2) {$Y_2$}
		
		(c1) edge[red] (a1)
		(c1) edge[red, bend right] (y1)
		(c2) edge[red] (a2)
		(c2) edge[red, bend left] (y2)
		
		
		(a1) edge[blue] (y1)
		(a1) edge[blue] (y2)
		(a2) edge[blue] (y1)
		(a2) edge[blue] (y2)
		
		node[below of=y1, yshift=0.2cm, xshift=0.5cm] (l) {$(b)$}
		;
		\end{scope}
%
%
%
%
%
		
		\begin{scope}[xshift=5.0cm] 
		\path[->, very thick]
		node[format] (a1) {$A_1$}
		node[format, above of=a1] (c1) {$C_1$}
		node[format, below of=a1] (m1) {$M_1$}
		node[format, below of=m1] (y1) {$Y_1$}
		
		node[format, right of=a1] (a2) {$A_2$}
		node[format, above of=a2] (c2) {$C_2$}
		node[format, below of=a2] (m2) {$M_2$}
		node[format, below of=m2] (y2) {$Y_2$}
		
		
		(c1) edge[blue] (a1)
		(c1) edge[blue, bend right=40] (m1)
		(c1) edge[blue, bend right=55] (y1)
		(a1) edge[blue] (m1)
		(m1) edge[blue] (y1)
		(a1) edge[<->, red, bend right=35] (y1)
		
		(c2) edge[blue] (a2)
		(c2) edge[blue, bend left=40] (m2)
		(c2) edge[blue, bend left=55] (y2)
		(a2) edge[blue] (m2)
		(m2) edge[blue] (y2)
		(a2) edge[<->, red, bend left=35] (y2)
		
		(a1) edge[blue] (y2)
		(a2) edge[blue] (y1)
		
		
		(a1) edge[blue] (m2)
		(a2) edge[blue] (m1)
		
		(m1) edge[-, brown] (m2)


		node[below of=y1, yshift=0.2cm, xshift=0.5cm] (l) {$(c)$}
		;
		\end{scope}

		\begin{scope}[xshift=8.6cm] 
		\path[->, very thick]
		node[] (a2) {} 
		node[format, above of=a2] (c2) {$C_2$}
		node[format, below of=a2] (m2) {$M_2$}
		node[format, below of=m2] (y2) {$Y_2$}
		
		node[left of=a2] (a1) {}
		node[format, left of=c2] (c1) {$C_1$}
		node[format, below of=a1] (m1) {$M_1$}
		
		(c1) edge[blue, bend right=40] (m1)
		
		(c2) edge[blue, bend left=40] (m2)
		(c2) edge[blue, bend left=55] (y2)
		(m2) edge[blue] (y2)
		
		
		

		
		(m1) edge[-, brown] (m2)
		node[below of=y2, yshift=0.2cm, xshift=-0.5cm] (l) {$(d)$}
		;
		\end{scope}
		
		\begin{scope}[xshift=10.6cm]
		\path[->, very thick]
		node[format] (a1) {$A_1$}
		node[format, above of=a1] (c1) {$C_1$}
		node[square, below of=a1] (m1) {$m_1$}
		node[format, below of=m1] (y1) {$Y_1$}
		
		node[format, right of=a1] (a2) {$A_2$}
		node[format, above of=a2] (c2) {$C_2$}
		node[square, below of=a2] (m2) {$m_2$}
		node[format, below of=m2] (y2) {$Y_2$}
		
		(c1) edge[blue] (a1)
		(c1) edge[blue, bend right=55] (y1)
		(a1) edge[<->, red, bend right=35] (y1)
		
		(c2) edge[blue] (a2)
		(c2) edge[blue, bend left=55] (y2)
		(a2) edge[<->, red, bend left=35] (y2)
		
		(m1) edge[blue] (y1)
		(m2) edge[blue] (y2)
		
		(a1) edge[blue] (y2)
		(a2) edge[blue] (y1)


		node[below of=y1, yshift=0.2cm, xshift=0.5cm] (l) {$(e)$}
		;
		\end{scope}
		
		\end{tikzpicture}
	\end{center}
	\caption{
		(a) A causal model representing the effect of community membership on article sharing, mediated by social network structure.
		(b) A causal model on dyads which is a variation of causal models of interference considered in \cite{ogburn14interference}.
		(c) A latent projection of the CG in (a) onto observed variables.
		(d) The graph representing $\mathcal{G}_{\mathbf{Y^*}}$ for the intervention operation $\text{do}(a_1)$ applied to (c).
		(e) The ADMG obtained by fixing $M_1, M_2$ in (c).
	}
	\label{fig:front_door}
\end{figure}
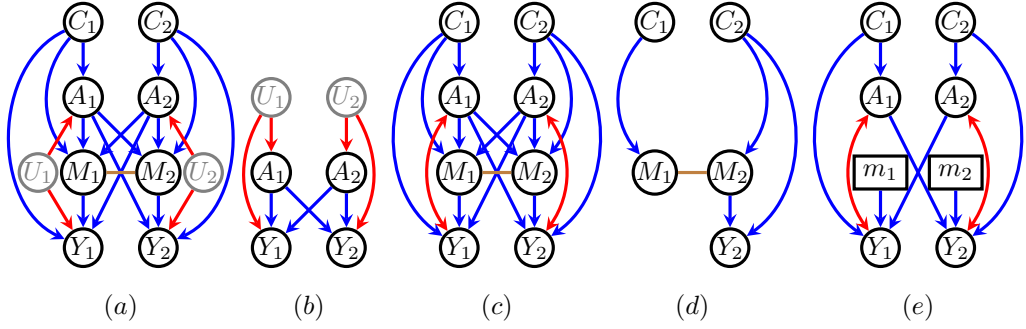

\section{Background on Causal Inference And Interference Problems}
\label{sec:review}

\subsection{Graph Theory}
\label{subsec:graphs}

We will consider causal models represented by mixed graphs containing directed ($\to$), bidirected ($\leftrightarrow$) and undirected ($-$) edges.  Vertices in these graphs and their corresponding random variables will be used interchangeably, denoted by capital letters, e.g. $V$; values, or realizations, of vertices and variables will be denoted by lowercase letters, e.g. $v$; bold letters will denote sets of variables or values e.g. $\mathbf{V}$ or $\mathbf{v}$.  We will denote the state space of a variable $V$ or a set of variables ${\bf V}$ as
${\mathfrak X}_V$, and ${\mathfrak X}_{\bf V}$.
Unless stated otherwise, all graphs will be assumed to have a vertex set denoted by ${\bf V}$.
For a mixed graph $\mathcal{G}$ of the above type, we denote the standard genealogic sets for a variable $V \in \mathbf{V}$ as follows: parents $\pa_{\mathcal{G}}(V) \equiv \{ W \in \mathbf{V} | W \rightarrow V \}$,
children $\ch_{\mathcal{G}}(V) \equiv \{ W \in \mathbf{V} | V \rightarrow W \}$,
siblings $\sib_{\cal G}(V) \equiv \{ W \in {\bf V} | W \leftrightarrow V \}$,
neighbors $\nb_{\cal G}(V) \equiv \{ W \in {\bf V} | W - V \}$,
ancestors $\an_{\mathcal{G}}(V) \equiv \{ W \in \mathbf{V} | W \rightarrow \dots \rightarrow V \}$,
descendants $\de_{\mathcal{G}}(V) \equiv \{ W \in \mathbf{V} | V \rightarrow \dots \rightarrow W \}$,
and non-descendants $\nd_{\mathcal{G}}(V) \equiv \mathbf{V} \setminus \de_{\mathcal{G}}(V)$.
We define the \emph{anterior} of $V$, or $\ant_{\G}(V)$, to be the set of all vertices with a partially directed path (a path containing only $\rightarrow$ and $-$ edges such that no $-$ edge can be oriented to induce a directed cycle) into $V$.
These relations generalize disjunctively to sets, for instance for a set ${\bf S}$, $\pa_{\cal G}({\bf S}) = \bigcup_{S \in {\bf S}} \pa_{\cal G}(S)$.  We also define the set $\pa_{\cal G}^s({\bf S})$ as $\pa_{\cal G}({\bf S}) \setminus {\bf S}$.
Given a graph ${\cal G}$ and a subset ${\bf S}$ of ${\bf V}$, define the \emph{induced subgraph} ${\cal G}_{\bf S}$ to be a graph with a vertex set ${\bf S}$ and all edges in ${\cal G}$ between elements in ${\bf S}$.

Given a mixed graph ${\cal G}$, we define a \emph{district} ${\bf D}$ to be a maximal set of vertices, where every vertex pair in ${\cal G}_{\bf D}$ is connected by a bidirected path (a path containing only $\leftrightarrow$ edges).  Similarly we define a \emph{block} ${\bf B}$ to be a maximal set of vertices, where every vertex pair in ${\cal G}_{\bf B}$ is connected by an undirected path (a path containing only $-$ edges).  Any block of size at least $2$ is called a non-trivial block.
We define a \emph{maximal clique} as a maximal set of vertices pairwise connected by undirected edges.
The set of districts in ${\cal G}$ is denoted by ${\cal D}({\cal G})$, the set of blocks is denoted by ${\cal B}({\cal G})$, the set non-trivial blocks is denoted by ${\cal B}^{nt}({\cal G})$, and the set of cliques is denoted by ${\cal C}({\cal G})$.
The district of $V$ is denoted by $\dis_{\G}(V)$.  By convention, for any $V$, $\dis_{\G}(V) \cap \de_{\G}(V) \cap \an_{\G}(V) \cap \ant_{\G}(V) = \{ V \}$.

A mixed graph is called \emph{segregated (SG)} if it contains no partially directed cycles, and no vertex has both neighbors and siblings, Fig.~\ref{fig:front_door} (c) is an example.
In a SG ${\cal G}$, ${\cal D}({\cal G})$ and ${\cal B}^{nt}({\cal G})$ partition ${\bf V}$.  A SG without bidirected edges is called a chain graph (CG) \cite{lauritzen96graphical}.  A SG without undirected edges is called an acyclic directed mixed graph (ADMG) \cite{richardson03markov}.  A CG without undirected edges or an ADMG without bidirected edges is a directed acyclic graph (DAG) \cite{pearl88probabilistic}.  A CG without directed edges is called an undirected graph (UG).  Given a CG ${\cal G}$, the augmented graph ${\cal G}^a$ is the UG where any adjacent vertices in ${\cal G}$ or any
elements in $\pa_{\cal G}({\bf B})$ for any ${\bf B} \in {\cal B}({\cal G})$ are connected by an undirected edge.


\subsection{Graphical Models}
\label{subsec:graphical}

A graphical model is 
a set of distributions with conditional independences represented by structures in a graph.  The following (standard) definitions appear in \cite{lauritzen96graphical}.  A DAG model, or a Bayesian network, is a set of distributions associated with a DAG ${\cal G}$ that can be written in terms of a DAG factorization: $p({\bf V}) = \prod_{V \in {\bf V}} p(V | \pa_{\cal G}(V))$.  A UG model, or a Markov random field, is a set of distributions associated with a UG ${\cal G}$ that can be written in terms of a UG factorization:
$p({\bf V}) = Z^{-1} \prod_{{\bf C} \in {\cal C}({\cal G})} \psi_{\bf C}({\bf C})$, where $Z$ is a normalizing constant.
A CG model is a set of distributions associated with a CG ${\cal G}$ that can be written in terms of the following two level factorization:
$p({\bf V}) = \prod_{{\bf B} \in {\cal B}({\cal G})} p({\bf B} | \pa_{\cal G}({\bf B}))$, where for each ${\bf B} \in {\cal B}({\cal G})$,
$p({\bf B} | \pa_{\cal G}({\bf B})) = Z(\pa_{\cal G}({\bf B}))^{-1} \prod_{{\bf C} \in {\cal C}(({\cal G}_{{\bf B} \cup \pa_{\cal G}({\bf B})})^a); {\bf C}
\not\subseteq \pa_{\cal G}({\bf B})} \psi_{\bf C}({\bf C})$.



\subsection{Causal Inference and Causal Models}
\label{subsec:causal}

A causal model of a DAG is also a set of distributions, but on counterfactual random variables.  Given $Y \in {\bf V}$ and ${\bf A} \subseteq {\bf V} \setminus \{ Y \}$, a counterfactual variable, or `potential outcome', written as $Y({\bf a})$, represents the value of $Y$ in a hypothetical situation where a set of \emph{treatments} ${\bf A}$ is set to values ${\bf a}$ by an \emph{intervention operation} \cite{pearl09causality}.  Given a set ${\bf Y}$, define ${\bf Y}({\bf a}) \equiv \{ {\bf Y} \}({\bf a}) \equiv \{ Y({\bf a}) \mid Y \in {\bf Y} \}$.  The distribution $p({\bf Y}({\bf a}))$ is sometimes written as $p({\bf Y} | \text{do}({\bf a}))$ \cite{pearl09causality}.


Causal models of a DAG ${\cal G}$ consist of distributions defined on counterfactual random variables
of the form $V({\bf a})$ where ${\bf a}$ are values of $\pa_{\cal G}(V)$.
In this paper we assume Pearl's functional model for a DAG $\mathcal{G}$ with vertices $\mathbf{V}$, where $V({\bf a})$ are determined by \emph{structural equations} $f_V({\bf a}, \epsilon_V)$, which remain invariant under any possible intervention on ${\bf a}$, with $\epsilon_V$ an exogenous disturbance variable which introduces randomness into $V$ even after all elements of $\pa_{\cal G}(V)$ are fixed.  Under Pearl's model, the distribution $p(\{ \epsilon_V | V \in {\bf V} \})$ is assumed to factorize as $\prod_{V \in {\bf V}} p(\epsilon_V)$.
This implies that the sets of variables $\{ \{V(\mathbf{a}_V) \mid \mathbf{a}_V \in \mathfrak{X}_{\pa_{\mathcal{G}}(V)} \} \mid V \in \mathbf{V} \}$ are mutually independent \cite{pearl09causality}.
The \emph{atomic counterfactuals} in the above set model the relationship between $\pa_{\cal G}(V)$, representing direct causes of $V$, and $V$ itself.  From these, all other counterfactuals may be defined using recursive substitution.  For any ${\bf A} \subseteq {\bf V} \setminus \{ V \}$,
$V({\bf a}) \equiv V({\bf a}_{\pa_{\cal G}(V) \cap {\bf A}},
\{ \pa_{\cal G}(V) \setminus {\bf A} \}({\bf a}))$.
For example, in the DAG in Fig.~\ref{fig:front_door} (b), $Y_1(a_1)$ is defined to be $Y_1(a_1, U_1, A_2(U_2))$.
Counterfactual responses to interventions are often compared on the mean difference scale for two values $a,a'$, representing cases and controls: $\E[Y(a)] - \E[Y(a')]$.  This quantity is known as the average causal effect (ACE).

A causal parameter is said to be \emph{identified} in a causal model if it is a function of the observed data distribution $p({\bf V})$.
Otherwise the parameter is said to be \emph{non-identified}.
In any causal model of a DAG ${\cal G}$, all interventional distributions
$p({\bf V} \setminus {\bf A} | \text{do}({\bf a}))$
are identified by the \emph{g-formula} \cite{robins86new}:
{\small
\begin{align}
p({\bf V} \setminus {\bf A} | \text{do}({\bf a})) =
\!\!\!
\prod_{V \in {\bf V} \setminus {\bf A}}
\left.
\!\!\!
p(V | \pa_{\cal G}(V)) \right|_{{\bf A}={\bf a}}
\label{eqn:g}
\end{align}
}
Note that the g-formula may be viewed as a modified (or \emph{truncated}) DAG factorization, with terms corresponding to elements in ${\bf A}$ missing.

\subsection{Modeling Dependent Data}
\label{subsec:dep-cg}

So far, the causal and statistical models we have introduced assumed data generating process that produce independent samples.
To capture examples of the sort we introduced in section \ref{sec:example}, we must generalize these models.
Suppose we analyze data with $M$ blocks with $N$ units each.
It is not necessary to assume that blocks are equally sized for the kinds of
problems we consider, but we make this assumption to simplify our notation.
Denote the variable $Y$ for the $i$'th unit in block $j$ as $Y^j_{i}$.
For each block $j$, let ${\bf Y}^j \equiv ( Y_1^j, \ldots, Y_N^j )$, and let
${\bf Y} \equiv ( {\bf Y}^1, \ldots, {\bf Y}^M )$. In some cases we will not be concerned with units' block memberships. In these cases we will accordingly omit the superscript and the subscript will index the unit with respect to all units in the network.



We are interested in counterfactual responses to interventions on ${\bf A}$,
treatments on all units in all blocks.  For any ${\bf a} \in {\mathfrak X}_{\bf A}$,
define $Y_i^j({\bf a})$ to be the potential response of unit $i$ in block $j$
to a hypothetical treatment assignment of ${\bf a}$ to ${\bf A}$.
We define ${\bf Y}^j({\bf a})$ and ${\bf Y}({\bf a})$ in the natural
way as vectors of responses, given a hypothetical treatment assignment to ${\bf a}$,
either for units in block $j$ or for all units, respectively.
Let ${\bf a}^{(j)}$ be a vector of values of ${\bf A}$,
where values assigned to units in block $j$ are \emph{free variables}, and
other values are \emph{bound variables}.  Furthermore, for any
$\tilde{\bf a}^j \in {\mathfrak X}_{{\bf A}^j}$, let ${\bf a}^{(j)}[\tilde{\bf a}^j]$ be
a vector of values which agrees on all bound values with ${\bf a}^{(j)}$,
but which assigns $\tilde{\bf a}^j$ to all units in block $j$ (e.g. which binds
free variables in ${\bf a}^{(j)}$ to $\tilde{\bf a}^j$).

A commonly made assumption is \emph{interblock non-interference},
also known as \emph{partial interference} in \cite{sobel06what,tchetgen12on},
where for any block $j$, treatments assigned to units in a block other than
$j$ do not affect the responses of any unit in block $j$.  Formally, this is
stated as
$(\forall j,{\bf a}^{(j)},{\bf a}'^{(j)},\tilde{\bf a}^j),
{\bf Y}^j({\bf a}^{(j)}[\tilde{\bf a}^{j}]) = 
{\bf Y}^j({\bf a}'^{(j)}[\tilde{\bf a}^{j}])$.
Counterfactuals under this assumption are written in a way that emphasizes they only depend
on treatments assigned within that block.
That is, for any ${\bf a}^{(j)}$, ${\bf Y}^j({\bf a}^{(j)}[\tilde{\bf a}^{j}]) \equiv {\bf Y}^j(\tilde{\bf a}^j)$.

In this paper we largely follow the convention of \cite{ogburn14interference}, where variables corresponding
to distinct units within a block are shown as distinct vertices in a graph.  As an example, Fig. \ref{fig:front_door} (b) represents
a causal model with observed data on multiple realizations of \emph{dyads} or blocks of two dependent units \cite{kenny06dyadic}.
Note that the arrow from $A_2$ to $Y_1$ in this model indicates that the treatment of unit $2$ in a block influences the outcome of
unit $1$, and similarly for treatment of unit $1$ and outcome of unit $2$.
In this model, a variation of models considered in \cite{ogburn14interference}, the interventional distributions
$p(Y_2 | \text{do}(a_1)) = p(Y_2 | a_1)$ and $p(Y_1 | \text{do}(a_2)) = p(Y_1 | a_2)$
even if $U_1,U_2$ are unobserved.


\section{Causal Inference with Hidden Variables}
\label{sec:hidden}

If a causal model contains hidden variables, only data on the observed marginal distribution is available.  In this case, not every interventional distribution is identified, and identification theory becomes more complex.  However, just as identified interventional distributions were expressible as a truncated DAG factorization via the g-formula (\ref{eqn:g}) in fully observed causal models, identified interventional distributions are expressible as a truncated \emph{nested} factorization \cite{richardson17nested} of a \emph{latent projection} ADMG \cite{verma90equiv} that represents a class of hidden variable DAGs that share identification theory.
In this section we define latent projection ADMGs, introduce the nested factorization with respect to an ADMG in terms of a fixing operator,
and re-express the ID algorithm \cite{tian02on,shpitser06id} as a truncated nested factorization.

\subsection{Latent Projection ADMGs}
\label{subsec:latent}

Given a DAG ${\cal G}({\bf V}\cup{\bf H})$, where ${\bf V}$ are observed and ${\bf H}$ are hidden variables, a {latent projection} ${\cal G}({\bf V})$ is the following ADMG with a vertex set ${\bf V}$. An edge $A \to B$ exists in ${\cal G}({\bf V})$ if there exists a directed path from $A$ to $B$ in ${\cal G}({\bf V}\cup{\bf H})$ with all intermediate vertices in ${\bf H}$.  Similarly, an edge $A \leftrightarrow B$ exists in ${\cal G}({\bf V})$ if there exists a path without consecutive edges $\to \circ \gets$ from $A$ to $B$ with the first edge on the path of the form $A \gets$ and the last edge on the path of the form $\to B$, and all intermediate vertices on the path in ${\bf H}$.
As an example of this operation, the graph in Fig.~\ref{fig:front_door} (c) is the latent projection of Fig.~\ref{fig:front_door} (a). Note that a variable pair in a latent projection ${\cal G}({\bf V})$ may be connected by both a directed and a bidirected edge, and that multiple distinct hidden variable DAGs ${\cal G}_1({\bf V} \cup {\bf H}_1)$ and ${\cal G}_2({\bf V} \cup {\bf H}_2)$ may share the same latent projection ADMG.

\subsection{The Nested Factorization}
\label{subsec:nested}

The nested factorization of $p({\bf V})$ with respect to an ADMG ${\G}({\bf V})$ is defined on \emph{kernel} objects derived from $p({\bf V})$ and \emph{conditional ADMGs} derived from $\G({\bf V})$.  The derivations are via a fixing operation, which can be causally interpreted as a single application of the g-formula on a single variable (to either a graph or a kernel) to obtain another graph or another kernel.


\subsubsection{Conditional Graphs And Kernels}
\label{subsubsec:cond}


A \textit{kernel} $q_{\mathbf{V}}(\mathbf{V} | \mathbf{W})$ is a mapping from values in $\mathbf{W}$ to normalized densities over $\mathbf{V}$ \cite{lauritzen96graphical}.  In other words, kernels act like conditional distributions in the sense that $\sum_{\mathbf{v} \in \mathbf{V}} q_{\mathbf{V}}(\mathbf{v} | \mathbf{w}) = 1, \forall \mathbf{w} \in \mathbf{W}$. Conditioning and marginalization in kernels are defined in the usual way.  For $\mathbf{A} \subseteq \mathbf{V}$, we define $q(\mathbf{A} | \mathbf{W}) \equiv \sum_{\mathbf{V} \setminus \mathbf{A}} q(\mathbf{V} | \mathbf{W})$ and $q(\mathbf{V} \setminus \mathbf{A} | \mathbf{A}, \mathbf{W}) \equiv {q(\mathbf{V} | \mathbf{W})}/{q(\mathbf{A} | \mathbf{W})}$.

A conditional acyclic directed mixed graph (CADMG) $\mathcal{G}(\mathbf{V}, \mathbf{W})$ is an ADMG in which the nodes are partitioned into $\mathbf{W}$, representing \textit{fixed variables}, and $\mathbf{V}$, representing \textit{random variables}.  Variables in $\mathbf{W}$ have the property that only outgoing directed edges may be adjacent to them.  Genealogic relationships generalize from ADMGs to CADMGs without change.  Districts are defined to be subsets of ${\bf V}$ in a CADMG $\G$, e.g. no element of ${\bf W}$ is in any element of ${\cal D}(\G)$.

\subsubsection{Fixability and Fixing}
\label{subsubsec:fix}

A variable $V \in \mathbf{V}$ in a CADMG $\mathcal{G}$ is \textit{fixable} if $\de_{\G}(V) \cap \dis_{\G}(V) = \emptyset$. In other words, $V$ is fixable if paths $V \leftrightarrow \dots \leftrightarrow B$ and $V \rightarrow \dots \rightarrow B$ do not \emph{both} exist in $\mathcal{G}$ for any $B \in \mathbf{V} \setminus \{ V \}$.
Given a CADMG $\G({\bf V},{\bf W})$ and $V \in {\bf V}$ fixable in $\G$, the fixing operator $\phi_V(\G)$ yields a new CADMG 
$\mathcal{G}'(\mathbf{V} \setminus \{V\} | \mathbf{W} \cup \{V\})$, where all edges with arrowheads into $V$ are removed, and all other edges in $\G$ are kept.  Similarly, given a CADMG $\G({\bf V},{\bf W})$, a kernel $q_{\bf V}({\bf V} | {\bf W})$, and $V \in {\bf V}$ fixable in $\G$, the fixing operator $\phi_V(q_{\bf V}; \G)$ yields a new kernel $q_{\mathbf{V} \setminus \{V\}}'(\mathbf{V} \setminus \{V\} | \mathbf{W} \cup \{V\}) \equiv \frac{q_{\mathbf{V}}(\mathbf{V} | \mathbf{W})}{q_{\mathbf{V}}(V | \nd_{\mathcal{G}}(V), \mathbf{W})}$. Note that fixing
is a probabilistic operation in which we divide a kernel by a conditional kernel. In some cases this operates as a conditioning operation, in other cases as a marginalization operation, and in yet other cases, as neither, depending on the structure of the kernel being divided.

For a set $\mathbf{S} \subseteq \mathbf{V}$ in a CADMG $\mathcal{G}$, if all vertices in ${\bf S}$ can be ordered into a sequence $\sigma_{\bf S} = \langle S_1, S_2, \dots \rangle$ such that $S_1$ is fixable in $\mathcal{G}$, $S_2$ in $\phi_{S_1}(\mathcal{G})$, etc., $\mathbf{S}$ is said to be \emph{fixable} in $\G$, ${\bf V} \setminus {\bf S}$ is said to be \emph{reachable} in $\G$, and $\sigma_{\bf S}$ is said to be valid.  A reachable set ${\bf C}$ is said to be \emph{intrinsic} if ${\cal G}_{\bf C}$ has a single district.
We will define $\phi_{\sigma_{\bf S}}(\G)$ and $\phi_{\sigma_{\bf S}}(q; \G)$ via the usual function composition to yield operators that fix all elements in ${\bf S}$ in the order given by $\sigma_{\bf S}$.

The distribution $p({\bf V})$ is said to obey the nested factorization for an ADMG $\G$ if there exists a set of kernels
$\{ q_{\bf C}({\bf C} \mid \pa_{\cal G}({\bf C})) \mid {\bf C} \text{ is intrinsic in }{\cal G} \}$ such that
for every fixable ${\bf S}$, and any valid $\sigma_{\bf S}$,
$\phi_{\sigma_{\bf S}}(p({\bf V});\G) = \prod_{{\bf D} \in {\cal D}(\phi_{\sigma_{\bf S}}(\G))} q_{\bf D}({\bf D} | \pa^s_{\G}({\bf D}))$.
All valid fixing sequences for ${\bf S}$ yield the same CADMG $\G({\bf V} \setminus {\bf S}, {\bf S})$, and
if $p({\bf V})$ obeys the nested factorization for $\G$, all valid fixing sequences for ${\bf S}$ yield the same kernel.
As a result, for any valid sequence $\sigma$ for ${\bf S}$, we will redefine the operator $\phi_{\sigma}$, for both graphs and kernels, to be $\phi_{\bf S}$.  In addition, it can be shown \cite{richardson17nested} that the above kernel set is characterized as:
\[
\{ q_{\bf C}({\bf C} \mid \pa_{\cal G}({\bf C})) \mid {\bf C} \text{ is intrinsic in }{\cal G} \} =
\{ \phi_{{\bf V} \setminus {\bf C}}(p({\bf V});{\cal G}) \mid {\bf C} \text{ is intrinsic in }{\cal G} \}.
\]
Thus, we can re-express the above nested factorization as stating that for any fixable set ${\bf S}$, we have
$\phi_{\bf S}(p({\bf V}); \G) = \prod_{{\bf D} \in {\cal D}(\phi_{{\bf S}}(\G))} \phi_{{\bf V} \setminus {\bf D}}(p({\bf V}); \G)$.
Since fixing is defined on CADMGs and kernels, the definition of nested Markov models generalizes in a straightforward way to a kernel
$q({\bf V} | {\bf W})$ being in the nested Markov model for a CADMG $\G({\bf V},{\bf W})$.  This holds if for every ${\bf S}$ fixable in
$\G({\bf V},{\bf W})$,
$\phi_{\bf S}(q({\bf V} | {\bf W}); \G) = \prod_{{\bf D} \in {\cal D}(\phi_{{\bf S}}(\G))} \phi_{{\bf V} \setminus {\bf D}}(q({\bf V} | {\bf W}); \G)$.

An important result in \cite{richardson17nested} states that if $p({\bf V} \cup {\bf H})$ obeys the factorization for a DAG $\G$ with vertex set ${\bf V} \cup {\bf H}$, then $p({\bf V})$ obeys the nested factorization for the latent projection ADMG $\G({\bf V})$.

\subsection{Identification in Hidden Variable Causal DAGs}
\label{subsec:id}

For any disjoint subsets ${\bf Y},{\bf A}$ of ${\bf V}$ in a latent projection $\G({\bf V})$ representing a causal DAG $\G({\bf V}\cup{\bf H})$, define ${\bf Y}^* \equiv \an_{\G({\bf V})_{{\bf V} \setminus {\bf A}}}({\bf Y})$.
Then $p({\bf Y} | \text{do}({\bf a}))$ is identified in $\G$ if \emph{and only if} every set ${\bf D} \in {\cal D}(\G({\bf V})_{{\bf Y}^*})$ is reachable (in fact, intrinsic).  Moreover, if identification holds, we have \cite{richardson17nested}:
{\small
\begin{align}
p({\bf Y} | \text{do}({\bf a})) = \sum_{{\bf Y}^* \setminus {\bf Y}} \prod_{{\bf D} \in {\cal D}(\G({\bf V})_{{\bf Y}^*})} \phi_{{\bf V} \setminus {\bf D}}(p({\bf V}); \G({\bf V})) \vert_{{\bf A} = {\bf a}}.
\label{eqn:id}
\end{align}
}
In other words, $p({\bf Y} | \text{do}({\bf a}))$ is only identified if it can be expressed as a factorization, where every piece corresponds to a kernel associated with a set intrinsic in $\G({\bf V})$.  Moreover, no piece in this factorization contains elements of ${\bf A}$ as random variables, just as was the case in (\ref{eqn:g}).  In fact, (\ref{eqn:id}) provides a concise formulation of the ID algorithm \cite{tian02on,shpitser06id} in terms of the nested Markov model in which the observed distribution in the causal problem lies.  For a full proof, see \cite{richardson17nested}.


\section{Chain Graphs For Causal Inference With Dependent Data}
\label{sec:cg}

We generalize causal models to represent settings with data dependence, specifically to cases where variables may exhibit stable but symmetric relationships.  These may correspond to friendship ties in a social network, physical proximity, or rules of infectious disease spread.  These stand in contrast to causal relationships which are also stable, but asymmetric.  We represent settings with both of these kinds of relationships using causal CG  models under the Lauritzen-Wermuth-Freydenburg (LWF) interpretation.  Though there are alternative conceptions of chain graphs \cite{drton09discrete}, we concentrate on LWF CGs here.  This is because LWF CGs yield observed data distributions with smooth parameterizations.  In addition, LWF CGs yield Markov properties where each unit's friends (and direct causes) screen the unit from other units in the network.  This sort of independence is intuitively appealing in many network settings.  Extensions of our results to other CG models are likely possible, but we leave them to future work.

LWF CGs were given a {causal} interpretation in \cite{lauritzen02chain}.  In a causal CG, the distribution $p({\bf B} | \pa_{\cal G}({\bf B}))$ for each block ${\bf B}$ is determined via a computer program that implements a Gibbs sampler on variables $B \in {\bf B}$, where the conditional distribution $p(B | {\bf B} \setminus \{ B \}, \pa_{\G}({\bf B}))$ is determined via a structural equation of the form
$f_B({\bf B} \setminus \{ B \}, \pa_{\G}({\bf B}), \epsilon_B)$.  This interpretation of $p({\bf B} | \pa_{\cal G}({\bf B}))$ allows the implementation of a simple intervention operation $\text{do}(b)$. The operation sets $B$ to $b$ by replacing the line of the Gibbs sampler program
that assigns $B$ to the value returned by $f_B({\bf B} \setminus \{ B \}, \pa_{\G}({\bf B}), \epsilon_B)$ (given a new realization of $\epsilon_B$), with an assignment of $B$ to the value $b$.
It was shown \cite{lauritzen02chain} that in a causal CG model, for any disjoint ${\bf Y},{\bf A}$, $p({\bf Y} | \text{do}({\bf a}))$ is identified by the CG version of the g-formula (\ref{eqn:g}):
$p({\bf Y} | \text{do}({\bf a})) = \prod_{\mathbf{B} \in \mathcal{B}(\mathcal{G})} p(\mathbf{B} \setminus \mathbf{A} | \pa(\mathbf{B}), \mathbf{B} \cap \mathbf{A})|_{\mathbf{A} = \mathbf{a}}$.

In our example above, stable symmetric relationships inducing data dependence, represented by undirected edges, coexist with hidden variables.  To represent causal inference in this setting, we generalize earlier developments for hidden variable causal DAG models to hidden variable causal CG models.  Specifically, we first define a latent projection analogue called the segregated projection for a large class of hidden variable CGs using segregated graphs (SGs).  We then define a factorization for SGs that generalizes the nested factorization and the CG factorization, and show that if a distribution $p({\bf V}\cup{\bf H})$ factorizes given a CG $\G({\bf V}\cup{\bf H})$ in the class, then $p({\bf V})$ factorizes according to the segregated projection $\G({\bf V})$.  Finally, we derive identification theory for hidden variable CGs as a generalization of (\ref{eqn:id}) that can be viewed as a truncated SG factorization.

\subsection{Segregated Projections Of Latent Variable Chain Graphs}
\label{subsec:sg-proj}

Fix a chain graph CG $\mathcal{G}$ and a vertex set $\mathbf{H}$ such that for all $H \in \mathbf{H}$, $H$ does not lie in ${\bf B} \cup \pa_{\mathcal{G}}(\mathbf{B})$, for any $\mathbf{B} \in \mathcal{B}^{nt}(\mathcal{G})$.  We call such a set $\mathbf{H}$ \emph{block-safe}.
\begin{dfn}
Given a CG $\mathcal{G}(\mathbf{V} \cup \mathbf{H})$ and a block-safe set $\mathbf{H}$, define a segregated projection graph $\mathcal{G}(\mathbf{V})$ with a vertex set $\mathbf{V}$.  Moreover, for any collider-free path from any two elements $V_1, V_2$ in $\mathbf{V}$, where all intermediate vertices are in $\mathbf{H}$, $\mathcal{G}(\mathbf{V})$ contains an edge with end points matching the path. That is, we have
    $V_1 \gets \circ \ldots \circ \to V_2$  leads to the edge $V_1 \leftrightarrow V_2$,
    $V_1 \to \circ \ldots \circ \to V_2$  leads to the edge $V_1 \to V_2$, and
    in  $\mathcal{G}(\mathbf{V})$.
\end{dfn}
As an example, the SG in Fig.~\ref{fig:front_door} (c) is a segregated projection of the hidden variable CG in Fig.~\ref{fig:front_door} (a).
While segregated graphs preserve conditional independence structure on the observed marginal of a CG for \emph{any} {\bf H} \cite{shpitser15segregated}, we chose to further restrict the set ${\bf H}$ in order to ensure that the directed edges in the segregated projection retain an intuitive causal interpretation of edges in a latent projection \cite{verma90equiv}.  That is, whenever $A \to B$ in a segregated projection, $A$ is a causal ancestor of $B$ in the underlying causal CG.
SGs represent latent variable CGs, meaning that they allow causal systems that model feedback that leads to network structures, of the sort considered in \cite{lauritzen02chain}, but simultaneously allow certain forms of unobserved confounding in such causal systems.

\subsection{Segregated Factorization}
\label{subseq:sg-m}

The segregated factorization of an SG can be defined as a product of two kernels which themselves factorize, one in terms of a CADMG (a conditional graph with only directed and bidirected arrows), and another in terms of a \emph{conditional chain graph (CCG)} $\mathcal{G}(\mathbf{V}, \mathbf{W})$, a CG with the property that the only type of edge adjacent to any element $W$ of $\mathbf{W}$ is a directed edge out of $W$.
A kernel $q(\mathbf{V} | \mathbf{W})$ is said to be Markov relative to the CCG
$\mathcal{G}(\mathbf{V}, \mathbf{W})$ if
    $q(\mathbf{V} | \mathbf{W}) = {Z(\mathbf{W})}^{-1} \prod_{{\bf B} \in {\cal B}(\G)} q({\bf B} | \pa_{\G}({\bf B}))$,
and
$q({\bf B} | \pa_{\cal G}({\bf B})) = Z(\pa_{\cal G}({\bf B}))^{-1} \prod_{{\bf C} \in {\cal C}(({\cal G}_{{\bf B} \cup \pa_{\cal G}({\bf B})})^a); {\bf C}
\not\subseteq \pa_{\cal G}({\bf B})} \psi_{\bf C}({\bf C})$,
for each ${\bf B} \in {\cal B}({\cal G})$.

We now show, given $p({\bf V})$ and an SG $\G({\bf V})$, how to construct the appropriate CADMG and CCG, and the two corresponding kernels.  Given a SG $\G$, let \emph{district variables} ${\bf D}^*$ be defined as $\bigcup_{{\bf D} \in {\cal D}(\G)} {\bf D}$, and
let \emph{block variables} ${\bf B}^*$ be defined as $\bigcup_{{\bf B} \in {\cal B}^{nt}(\G)} {\bf B}$.  Since ${\cal D}(\G)$ and ${\cal B}^{nt}(\G)$ partition ${\bf V}$ in a SG, ${\bf B}^*$ and ${\bf D}^*$ partition ${\bf V}$ as well.
Let the induced CADMG $\mathcal{G}^d$ of a SG $\mathcal{G}$ be the graph containing the vertex sets $\mathbf{D}^*$ as $\mathbf{V}$ and $\pa^s_{\mathcal{G}}(\mathbf{D}^*)$ as $\mathbf{W}$, and which inherits all edges in $\mathcal{G}$ between ${\bf D}^*$, and all directed edges from $\pa^s_{\mathcal{G}}(\mathbf{D}^*)$ to ${\bf D}^*$ in $\G$.
Similarly, let the induced CCG $\mathcal{G}^b$ of $\mathcal{G}$ be the graph containing the vertex set $\mathbf{B}^*$ as $\mathbf{V}$ and $\pa^s_{\mathcal{G}}(\mathbf{B}^*)$ as $\mathbf{W}$, and which inherits all edges in $\mathcal{G}$ between ${\bf B}^*$, and all
directed edges from $\pa_{\mathcal{G}}(\mathbf{B}^*)$ to ${\bf B}^*$.
We say that $p({\bf V})$ obeys the factorization of a SG $\G({\bf V})$ if $p({\bf V}) = q({\bf D}^* | \pa^s_{\G}({\bf D}^*)) q({\bf B}^* | \pa_{\G}({\bf B}^*))$, $q({\bf B}^* | \pa_{\G}({\bf B}^*))$ is Markov relative to the CCG $\G^b$, and $q({\bf D}^* | \pa^s_{\G}({\bf D}^*))$ is in the nested Markov model of the CADMG $\G^d$.

The following theorem gives the relationship between a joint distribution that factorizes given a hidden variable CG $\G$, its marginal distribution, and the corresponding segregated factorization.  This theorem is a generalization of the result proven in \cite{richardson17nested} relating hidden variable DAGs and latent projection ADMGs.  The proof is deferred to the appendix.
\begin{thm}
\label{thm:sg-f}
If $p(\mathbf{V} \cup \mathbf{H})$ obeys the CG factorization relative to $\mathcal{G}(\mathbf{V} \cup \mathbf{H})$, and $\mathbf{H}$ is block-safe then $p(\mathbf{V})$ obeys the segregated factorization relative to the segregated projection $\mathcal{G}({\bf V})$.
\end{thm}

\section{A Complete Identification Algorithm for Latent Variable Chain Graphs}
\label{sec:sg-id}
With Theorem \ref{thm:sg-f} in hand, we are ready to characterize general non-parametric identification of interventional distributions in hidden variable causal chain graph models, where hidden variables form a block-safe set.  This result can be viewed on the one hand as a generalization of the CG g-formula derived in \cite{lauritzen02chain}, and on the other hand as a generalization of the ID algorithm (\ref{eqn:id}).
\begin{thm}
\label{thm:sg-id}
Assume $\mathcal{G}(\mathbf{V} \cup \mathbf{H})$ is a causal CG, where ${\bf H}$ is block-safe. Fix disjoint subsets $\mathbf{Y},\mathbf{A}$ of $\mathbf{V}$. Let $\mathbf{Y}^* = \ant_{\mathcal{G}(\mathbf{V})_{\mathbf{V} \setminus \mathbf{A}}} \mathbf{Y}$. Then $p(\mathbf{Y} | \text{do}(\mathbf{a}))$ is identified from $p(\mathbf{V})$
if and only if every element in $\mathcal{D}(\widetilde{\mathcal{G}}^d)$ is reachable in ${\mathcal{G}}^d$, where 
$\widetilde{\mathcal{G}}^d$ is the induced CADMG of $\G({\bf V})_{{\bf Y}^*}$.

Moreover, if $p(\mathbf{Y} | \text{do}(\mathbf{a}))$ is identified, it is equal to
{\small
\begin{align*}
   \sum_{\mathbf{Y}^* \setminus \mathbf{Y}} & \left[ \prod_{\mathbf{D} \in \mathcal{D}(\widetilde{\mathcal{G}}^d)} \phi_{\mathbf{D}^* \setminus \mathbf{D}}(q(\mathbf{D}^* | \pa_{\mathcal{G}({\bf V})}(\mathbf{D}^*)); \mathcal{G}^d) \right] 
    \left[ \left.\prod_{\mathbf{B} \in \mathcal{B}(\widetilde{\mathcal{G}}^b)} p(\mathbf{B} \setminus \mathbf{A} | \pa_{\mathcal{G}({\bf V})_{\mathbf{Y}^*}}(\mathbf{B}), \mathbf{B} \cap \mathbf{A}) \right] \right|_{\mathbf{A} = \mathbf{a}}
\end{align*}
}
where
$q(\mathbf{D}^* | \pa_{\mathcal{G}({\bf V})}(\mathbf{D}^*)) =
p({\bf V}) / (\prod_{{\bf B} \in {\cal B}^{nt}(\G({\bf V}))} p({\bf B} | \pa_{\G({\bf V})}({\bf B}))$, and
$\widetilde{\mathcal{G}}^b$ is the induced CCG of $\G({\bf V})_{{\bf Y}^*}$.
\end{thm}

To illustrate the application of this theorem, consider the SG $\G$ in Fig.~\ref{fig:front_door} (c), where we are interested in
$p(Y_2 | \text{do}(a_1,a_2))$.  It is easy to see that $\mathbf{Y^*} = \{C_1, C_2, M_1, M_2, Y_2\}$ (see $\mathcal{G}_{\mathbf{Y^*}}$ in Fig. \ref{fig:front_door} (d)) with $\mathcal{B}(\G_{{\bf Y}^*}) = \{\{M_1, M_2\}\}$ and $\mathcal{D}(\G_{{\bf Y}^*}) = \{\{C_1\}, \{C_2\}, \{Y_2\}\}$. 
The chain graph factor of the factorization in Theorem \ref{thm:sg-id} is $p(M_1, M_2 | A_1 = a_1, A_2, C_1, C_2)$. Note that this expression further factorizes according to the (second level) undirected factorization of blocks in a CCG. For the three district factors
$\{ C_1 \},\{ C_2 \},\{Y_2\}$ in
Fig.~\ref{fig:front_door} (d), we must fix variables in three different sets $\{C_2, A_1, A_2, Y_1, Y_2\}$, $\{C_1, A_1, A_2, Y_1, Y_2\}$,
$\{C_1, C_2, A_1, Y_1, A_2\}$ in $\mathcal{G}^d$, shown in Fig. \ref{fig:front_door} (e).
We defer the full derivation involving the fixing operator to the supplementary material.
The resulting identifying functional for $p(Y_2 | \text{do}(a_1,a_2))$ is:
{\small
\begin{equation}
\begin{split}
    \sum_{\{C_1, C_2, M_1, M_2\}} &
    p(M_1, M_2 | a_1, a_2,C_1,C_2) 
   \sum_{A_2} p(Y_2 | a_1, A_2, 
   	M_2, C_2)p(A_2|C_2) p(C_1) p(C_2)  
\end{split}\label{eqn:functional}
\end{equation}
}

\section{Experiments}
\label{sec:exp}
We now illustrate how identified functionals given by Theorem \ref{thm:sg-id} may be estimated from data.  Specifically we consider
network average effects (N.E.), the network analogue of the average causal effect (ACE), as defined in \cite{hudgens08toward}:
{\small
\begin{align*}
    \text{NE}^i(\mathbf{a}_{-i}) &= \frac{1}{N} \sum_i E[{Y}_i(A_i = 1, \mathbf{A}_{-1} = 1)] - E[{Y}_i(A_i = 0,\mathbf{A}_{-i} = 0)]
\end{align*}
}
in our article sharing example described in section \ref{sec:example}, and shown in simplified form (for two units) in Fig. \ref{fig:front_door} (a).  The experiments and results we present here generalize easily to other network effects such as direct and spillover effects \cite{hudgens08toward}, although we do not consider this here in the interests of space. For purposes of illustration we consider a simple setting where the social network is a $3$-regular graph, with networks of size $N = [400, 800, 1000, 2000]$.  Under the hidden variable CG model we described in section \ref{sec:example}, the above effect is identified by a functional which generalizes (\ref{eqn:functional}) from a network of size $2$ to a larger network.  Importantly, since we assume a single connected network of $M$ variables, we are in the \emph{full interference setting} where only a single sample from $p(M_1, \ldots M_N | A_1, \ldots, A_N, C_1, \ldots, C_N)$ is available.  This means that while the standard maximum likelihood plug-in estimation strategy is possible for models for $Y_i$ and $A_i$ in (\ref{eqn:functional}), the strategy does not work for the model for $M$.  Instead, we adapt the auto-g-computation approach based on the pseudo-likelihood and coding estimators proposed in \cite{tchetgen2017auto}, which is appropriate for full interference settings with a Markov property given by a CG, as part of our estimation procedure.  Note that the approach in \cite{tchetgen2017auto} was applied for a special case of the set of causal models considered here, in particular those with no unmeasured confounding.  Here we use the same approach for estimating general functionals in models that may include unobserved confounders between treatments and outcomes.  In fact, our example model is analogous to the model in \cite{tchetgen2017auto}, in the same way that the front-door criterion is to the backdoor criterion in causal inference under the assumption of iid data \cite{pearl09causality}.

Our detailed estimation strategy, along with a more detailed description of our results, is described in the appendix.
We performed $1000$ bootstrap samples of the $4$ different networks.
Since calculating the true causal effects is intractable even if true model parameters are known, we calculate the approximate `ground truth' for each intervention by sampling from our data generating process under the intervention $5$ times and averaging the relevant effect.  We calculated the (approximation of) the bias of each effect by subtracting the estimate from the `ground truth.'
The `ground truth' network average effects range from $-.453$ to $-.456$. As shown in Tables \ref{tab:bias_results_ci} and \ref{tab:bias_results}, both estimators recover the ground truth effect with relatively small bias.  Estimators for effects which used the pseudo-likelihood estimator for $M$ generally have lower variance than those that used the coding estimator for $M$, which is expected due to the greater efficiency of the former. This behavior was also observed in \cite{tchetgen2017auto}. In both estimators, bias decreases with network size.  This is also expected intuitively, although detailed asymptotic theory for statistical inference in networks is currently an open problem, due to dependence of samples.

\begin{table}[h]
	\begin{center}
		\begin{tabular}{|*6{p{18mm}|}}
			\hline
			\multicolumn{6}{|c|}{95\% Confidence Intervals of Bias of Network Average Effects}\\
			\hline
			& $N$ & 400 & 800 & 1000 & 2000\\
			\hline
			\multirow{2}{*}{Estimator} & Coding &(-.157, .103)&(-.129, .106)&(-.100, .065)&(-.086, .051)\\
			\cline{2-6}
			&Pseudo &(-.133, .080)&(-.099, .089)&(-.116, .074)&(-.070, .041)\\
			\hline
		\end{tabular}
	\end{center}
	\caption{95\% confidence intervals for the bias of each estimating method for the network average effects. All intervals cover the approximated ground truth since they include $0$}
	\label{tab:bias_results_ci}
\end{table}

\begin{table}[h]
	\begin{center}
		\begin{tabular}{|*6{p{18mm}|}}
			\hline
			\multicolumn{6}{|c|}{Bias of Network Average Effects}\\
			\hline
			& $N$ & 400 & 800 & 1000 & 2000\\
			\hline
			\multirow{2}{*}{Estimator} & Coding & -.000 (.060)& -.020 (.051) & -.024 (.052) & -.022 (.034)\\
			\cline{2-6}
			&Pseudo &.006 (.052)& -.023 (.042) & -.023 (.042) & -.021 (.026)\\
			\hline
		\end{tabular}
	\end{center}
	\caption{The biases of each estimating method for the network average effects. Standard deviation of the bias of each estimate is given in parentheses.}
	\label{tab:bias_results}
\end{table}


\section{Conclusion}
\label{sec:conclusion}
In this paper, we generalized existing non-parametric identification theory for hidden variable causal DAG models to hidden variable causal chain graph models, which can represent both causal relationships, and stable symmetric relationships that induce data dependence.  Specifically, we gave a representation of all identified interventional distributions in such models as a truncated factorization associated with \emph{segregated graphs}, mixed graphs containing directed, undirected, and bidirected edges which represent marginals of chain graphs.

We also demonstrated how statistical inference may be performed on identifiable causal parameters, by adapting a combination of maximum likelihood plug in estimation, and methods based on coding and pseudo-likelihood estimators that were adapted for full interference problems in \cite{tchetgen2017auto}.  We illustrated our approach with an example of calculating the effect of community membership on article sharing if the effect of the former on the latter is mediated by a complex social network of units inducing full dependence.


\section{Acknowledgements}
\label{sec:ack}
The second author would like to thank the American Institute of Mathematics for supporting this research via the SQuaRE program.
This project is sponsored in part by the National Institutes of Health grant R01 AI127271-01 A1, the
Office of Naval Research grant N00014-18-1-2760 and the Defense Advanced Research Projects Agency (DARPA) under contract HR0011-18-C-0049.  The content of the information does not necessarily reflect the position or the policy of the Government, and no official endorsement should be inferred.

\bibliographystyle{abbrv}
\bibliography{references}

\appendix

\section{Proofs}

\begin{thma}{\ref{thm:sg-f}}
If $p(\mathbf{V} \cup \mathbf{H})$ obeys the CG factorization relative to $\mathcal{G}(\mathbf{V} \cup \mathbf{H})$, and $\mathbf{H}$ is block-safe then $p(\mathbf{V})$ obeys the segregated factorization relative to the segregated projection $\mathcal{G}({\bf V})$.
\end{thma}

\begin{prf}
Assume the premise of the theorem. Then, $p(\mathbf{O} \cup \mathbf{H}) = \prod_{\mathbf{B} \in \mathcal{B}(\mathcal{G})} p(\mathbf{B} | \pa_{\mathcal{G}} (\mathbf{B}))$.

For every ${\bf D} \in {\cal D}(\G({\bf V}))$, let ${\bf H}_{\bf D} \equiv {\bf H} \cap \an_{{\G}_{{\bf D} \cup {\bf H}}}({\bf D})$.  Then
$p({\bf V})$ is equal to
\begin{align*}
& \sum_{\bf H}
\left( \prod_{{\bf B} \in {\cal B}^{nt}(\G)} p({\bf B} | \pa_{\G}({\bf B})) \right)
\left( \prod_{\{ B \} \not\in {\cal B}^{nt}(\G)} p(B | \pa_{\G}(B)) \right)\\
&= \left( \prod_{{\bf B} \in {\cal B}^{nt}(\G)} p({\bf B} | \pa_{\G}({\bf B})) \right)
\prod_{{\bf D} \in {\cal D}(\G({\bf V}))}
\sum_{{\bf H}_{\bf D}} \left( \prod_{B \in {\bf D}} p(B | \pa_{\G}(B)) \right)\\
&= \left( \prod_{{\bf B} \in {\cal B}^{nt}(\G)} p({\bf B} | \pa_{\G}({\bf B})) \right)
\prod_{{\bf D} \in {\cal D}(\G({\bf V}))}
q({\bf D} | \pa^s_{\G({\bf V})}({\bf D}))\\
&= q({\bf B}^* | \pa_{\G({\bf V})}({\bf B}^*))
q({\bf D}^* | \pa^s_{{\cal G}({\bf V})}({\bf D}^*)).
\end{align*}
The fact that $q({\bf B}^* | \pa_{\G({\bf V})}({\bf B}^*))$ factorizes according to the CCG $\G^b$ follows by construction.

Let $\widetilde{\bf B} \equiv \{ B \in {\bf V} \cup {\bf H} \mid \{ B \} \not\in{\cal B}^{nt}(\G) \}$.  Then
\begin{align*}
q(\widetilde{\bf B} | \pa^s_{\G}(\widetilde{\bf B})) =  \prod_{B : \{ B \} \not\in {\cal B}^{nt}(\G)} p(B | \pa_{\G}(B))
\end{align*}
factorizes according to the CADMG (in fact a conditional DAG) $\G(\widetilde{\bf B},\pa^s_{\G}(\widetilde{\bf B}))$ obtained from
$\G({\bf V}\cup{\bf H})$ by making all elements in $\pa^s_{\G}(\widetilde{\bf B})$ fixed, and all elements $\widetilde{\bf B}$ random,
keeping all edges among $\widetilde{\bf B}$ in $\G$, and all outgoing directed edges from $\pa^s_{\G}(\widetilde{\bf B})$ to
$\widetilde{\bf B}$ in $\G$.  The fact that $q({\bf D}^* | \pa_{\G({\bf V})}({\bf D}^*))$ factorizes according $\G^d$, the latent projection CADMG obtained from $\G(\widetilde{\bf B},\pa^s_{\G}(\widetilde{\bf B}))$ by treating ${\bf H}$ as hidden variables now follows by the inductive application of Lemmas 46 and 49 in \cite{richardson17nested} to $q(\widetilde{\bf B} | \pa^s_{\G}(\widetilde{\bf B}))$ and $\G(\widetilde{\bf B},\pa^s_{\G}(\widetilde{\bf B}))$.
\end{prf}

\begin{thma}{\ref{thm:sg-id}}
Assume $\mathcal{G}(\mathbf{V} \cup \mathbf{H})$ is a causal CG, where ${\bf H}$ is block-safe. Fix disjoint subsets $\mathbf{Y},\mathbf{A}$ of $\mathbf{V}$. Let $\mathbf{Y}^* = \ant_{\mathcal{G}(\mathbf{V})_{\mathbf{V} \setminus \mathbf{A}}} \mathbf{Y}$. Then $p(\mathbf{Y} | \text{do}(\mathbf{a}))$ is identified from $p(\mathbf{V})$
if and only if every element in $\mathcal{D}(\widetilde{\mathcal{G}}^d)$ is reachable in ${\mathcal{G}}^d$, where 
$\widetilde{\mathcal{G}}^d$ is the induced CADMG of $\G({\bf V})_{{\bf Y}^*}$.

Moreover, if $p(\mathbf{Y} | \text{do}(\mathbf{a}))$ is identified, it is equal to
\begin{align}
   \sum_{\mathbf{Y}^* \setminus \mathbf{Y}} & \left[ \prod_{\mathbf{D} \in \mathcal{D}(\widetilde{\mathcal{G}}^d)} \phi_{\mathbf{D}^* \setminus \mathbf{D}}(q(\mathbf{D}^* | \pa_{\mathcal{G}({\bf V})}(\mathbf{D}^*)); \mathcal{G}^d) \right] 
    \left[ \left.\prod_{\mathbf{B} \in \mathcal{B}(\widetilde{\mathcal{G}}^b)} p(\mathbf{B} \setminus \mathbf{A} | \pa_{\mathcal{G}({\bf V})_{\mathbf{Y}^*}}(\mathbf{B}), \mathbf{B} \cap \mathbf{A}) \right] \right|_{\mathbf{A} = \mathbf{a}}
    \label{eqn:seg_id}
\end{align}
where
$q(\mathbf{D}^* | \pa_{\mathcal{G}({\bf V})}(\mathbf{D}^*)) =
p({\bf V}) / (\prod_{{\bf B} \in {\cal B}^{nt}(\G({\bf V}))} p({\bf B} | \pa_{\G({\bf V})}({\bf B}))$, and
$\widetilde{\mathcal{G}}^d$ is the induced CCG of $\G({\bf V})_{{\bf Y}^*}$.
\end{thma}

\begin{prf} We proceed by proving a series of subclaims.


\textbf{Claim 1}: \textit{If $p(\mathbf{O})$ obeys the segregated factorization relative to $\mathcal{G}(\mathbf{O})$, then $p(\mathbf{A})$ obeys the segregated factorization relative to $\mathcal{G}(\mathbf{O})_{\mathbf{A}}$} for any subset ${\bf A} \subseteq {\bf O}$ anterial in ${\cal G}({\bf O})$.  A set ${\bf A}$ is anterial if, whenever $X \in {\bf A}$, $\ant_{\cal G}(X) \subseteq {\bf A}$.

We show this by induction.  Assume $p(\mathbf{O})$ obeys the segregated factorization relative to $\mathcal{G}(\mathbf{O})$, and
${\bf A}$ consists of all elements in ${\bf O}$ other than those in ${\bf B} \in {\cal B}^{nt}({\cal G}({\bf O}))$.  Then by writing
$p({\bf A}) = \sum_{{\bf B}} p({\bf O})$ as a segregated factorization for $p({\bf O})$, we note that the nested factorization remains unchanged by the marginalization, and the block factorization remains unchanged, except the factor corresponding to ${\bf B}$ is removed.

Similarly, assume $p(\mathbf{O})$ obeys the segregated factorization relative to $\mathcal{G}(\mathbf{O})$, and
${\bf A}$ consists of all elements in ${\bf O}$ other than some element $B$ not in any ${\bf B} \in {\cal B}^{nt}({\cal G}({\bf O}))$ such that
$\ch_{\cal G}(B)$ is empty.  Then by writing $p({\bf A}) = \sum_{{\bf B}} p({\bf O})$ as a segregated factorization for $p({\bf O})$, we note that the block factorization remains unchanged by the marginalization, and the kernel
\begin{align*}
q({\bf B}^* \setminus \{ B \} \mid \pa^s_{{\cal G}({\bf O})}({\bf B}^*)) = \sum_{B} \frac{p({\bf V})}{\prod_{{\bf B} \in {\cal B}^{nt}(\G({\bf V}))} p({\bf B} | \pa_{\G({\bf V})}({\bf B}))}
\end{align*}
is nested Markov relative to the CADMG $\tilde{\cal G}({\bf O})^d$ obtained from ${\cal G}({\bf O})^d$ by removing $B$ and all edges adjacent to $B$.  To see this, note that reachable sets in $\tilde{\cal G}({\bf O})^d$ are a strict subset of reachable sets in ${\cal G}({\bf O})^d$, since $B$ is fixable in ${\cal G}({\bf O})^d$, and moreover all kernels corresponding to reachable sets in $\tilde{\cal G}({\bf O})^d$ may be obtained from $q({\bf B}^* \mid \pa^s_{{\cal G}({\bf O})}({\bf B}^*))$ by marginalizing $B$ first, and applying the fixing operator to remaining variables in ${\cal B}^* \setminus \{ B \}$.  As a result, the nested global Markov property for the former graph is implied by the nested global Markov property of the latter graph, proving our claim.

\textbf{Claim 2}: \textit{The algorithm specified by the equation (\ref{eqn:seg_id}) is sound for identification of $p({\bf Y} | \text{do}({\bf a}))$.}

Per claim 1, without loss of generality assume ${\bf Y}$ has no children in ${\cal G}({\bf O})$. Consider the chain graph g-formula:
\begin{align*}
    p(\mathbf{Y}(\mathbf{a})) = \prod_{\mathbf{B} \in \mathcal{B}(\mathcal{G}(\mathbf{O} \cup \mathbf{H}))} p(\mathbf{B} \setminus \mathbf{A} | \pa_{\mathcal{G}}(\mathbf{B}), \mathbf{B} \cap \mathbf{A})|_{\mathbf{A} = \mathbf{a}}.
\end{align*}
We can decompose this into factors relating to the non-trivial blocks and districts in the graph:
\begin{align*}
    p(\mathbf{Y}(\mathbf{a})) &= \prod_{\mathbf{B} \in \mathcal{B}^{nt}(\mathcal{G}(\mathbf{O} \cup \mathbf{H}))} p(\mathbf{B} \setminus \mathbf{A} | \pa_{\mathcal{G}}(\mathbf{B}), \mathbf{B} \cap \mathbf{A})|_{\mathbf{A} = \mathbf{a}}\\
    \times &\prod_{\mathbf{D} \in \mathcal{D}(\mathcal{G}(\mathbf{O} \cup \mathbf{H}))} p(\mathbf{D} \setminus \mathbf{A} | \pa_{\mathcal{G}}(\mathbf{D}), \mathbf{D} \cap \mathbf{A})|_{\mathbf{A} = \mathbf{a}}.
\end{align*}
Since ${\bf H}$ is block-safe, the factors in the first term -- those that correspond to non-trivial blocks -- are the same in the segregated graph as in the original chain graph and thus we can re-write the above as:
\begin{align*}
    p(\mathbf{Y}(\mathbf{a})) &= \prod_{\mathbf{B} \in \mathcal{B}^{nt}(\mathcal{G}_{\mathbf{Y}^*})} p(\mathbf{B} \setminus \mathbf{A} | \pa_{\mathcal{G}}(\mathbf{B}), \mathbf{B} \cap \mathbf{A})|_{\mathbf{A} = \mathbf{a}}\\
    \times &\prod_{\mathbf{D} \in \mathcal{D}(\mathcal{G}(\mathbf{O} \cup \mathbf{H}))} p(\mathbf{D} \setminus \mathbf{A} | \pa_{\mathcal{G}}(\mathbf{D}), \mathbf{D} \cap \mathbf{A})|_{\mathbf{A} = \mathbf{a}}.
\end{align*}
Meanwhile the factors in the second term describe a kernel $q(\mathbf{D}^* | \pa_{\mathcal{G}_{(\mathbf{O} \cup \mathbf{H})}}(\mathbf{D}^*))$ associated with a CADG $\mathcal{G}(\mathbf{O} \cup \mathbf{H}, \mathbf{B}^*)$ which we can manipulate to obtain the desired result by following the argument in the proof of Theorem 60 in \cite{richardson17nested}. 

Let $\mathbf{A}^* = \mathbf{O} \setminus \mathbf{Y}^* \supseteq \mathbf{A}$. By the global Markov property of conditional DAGs (CDAGs) proven in \cite{richardson17nested}, $p(\mathbf{Y}^* | \textit{do}_{\mathcal{G}(\mathbf{O} \cup \mathbf{H}, \mathbf{B}^*)}(\mathbf{a})) = p(\mathbf{Y}^* | \textit{do}_{\mathcal{G}(\mathbf{O} \cup \mathbf{H}, \mathbf{B}^*)}(\mathbf{a}^*))$.

Let $\mathcal{G}^*((\mathbf{O} \setminus \mathbf{A}^*) \cup \mathbf{H}, \mathbf{B}^* \cup \mathbf{A}^*) = \phi_{\mathbf{A}^*}(\mathcal{G}(\mathbf{O} \cup \mathbf{H}, \mathbf{B}^*))$. Let $\sigma_{\mathbf{H}}$ denote the \emph{latent projection operation} such that $\sigma_{\mathbf{H}}(\mathcal{G}(\mathbf{O} \cup \mathbf{H}) = \mathcal{G}(\mathbf{O})$.
Then, by commutativity of $\sigma_{\mathbf{H}}$ and the fixing operator (Corollary 53 in \cite{richardson17nested}), $\sigma_{\mathbf{H}}(\phi_{\mathbf{A}^*}(\mathcal{G}(\mathbf{O} \cup \mathbf{H}, \mathbf{B}^*))) = \phi_{\mathbf{A}^*}(\sigma_{\mathbf{H}}(\mathcal{G}(\mathbf{O} \cup \mathbf{H}, \mathbf{B}^*))) = \mathcal{G}^*(\mathbf{Y}^*, \mathbf{B}^* \cup \mathbf{A}^*)$.
By definition of induced subgraphs, $\mathcal{G}(\mathbf{O}, \mathbf{B}^*)_{\mathbf{Y}^*} = (\phi_{\mathbf{A}^*}(\mathcal{G}(\mathbf{O}, \mathbf{B}^*)))_{\mathbf{Y}^*}$.
By these two equalities, we have $\mathcal{G}(\mathbf{O}, \mathbf{B}^*)_{\mathbf{Y}^*} = \mathcal{G}^*(\mathbf{O}, \mathbf{B}^* \cup \mathbf{A}^*)_{\mathbf{Y}^*}$ and thus $\mathcal{D}(\mathcal{G}(\mathbf{O}, \mathbf{B}^*)_{\mathbf{Y}^*}) = \mathcal{D}(\mathcal{G}^*(\mathbf{Y}^*, \mathbf{B}^* \cup \mathbf{A}^*))$.

For each $\mathbf{D} \in \mathcal{D}(\mathcal{G}^*(\mathbf{Y}^*, \mathbf{B}^* \cup \mathbf{A}^*))$, let $\mathbf{H}_{\mathbf{D}} \equiv \mathbf{H} \cap \an_{\mathcal{G}(\mathbf{O} \cup \mathbf{H}, \mathbf{B}^*)_{\mathbf{D} \cup \mathbf{H}}}(\mathbf{D})$ and $\mathbf{H}^* \equiv \bigcup_{\mathbf{D} \in \mathcal{D}(\mathcal{G}^*(\mathbf{Y}^*, \mathbf{B}^* \cup \mathbf{A}^*))} \mathbf{H}_{\mathbf{D}}$.
Then, by construction, if $\mathbf{D}, \mathbf{D}' \in \mathcal{D}(\mathcal{G}^*(\mathbf{Y}^*, \mathbf{B}^* \cup \mathbf{A}^*)$ and $\mathbf{D} \neq \mathbf{D}'$ then $\mathbf{H}_{\mathbf{D}} \cap \mathbf{H}_{\mathbf{D}'} = \emptyset$. Additionally, for all $\mathbf{D} \in \mathcal{D}(\mathcal{G}^*(\mathbf{Y}^*, \mathbf{B}^* \cup \mathbf{A}^*)$, it is the case that $\pa_{\mathcal{G}(\mathbf{O} \cup \mathbf{H}, \mathbf{B}^*)}(\mathbf{D} \cup \mathbf{H}_{\mathbf{D}}) \cap \mathbf{H}^* = \mathbf{H}_{\mathbf{D}}$. And $\mathbf{Y}^* \cup \mathbf{H}^*$ is ancestral in $\mathcal{G}(\mathbf{O} \cup \mathbf{H}, \mathbf{B}^*)$ which implies that if $v \in \mathbf{Y}^* \cup \mathbf{H}^*$, then $\pa_{\mathcal{G}(\mathbf{O} \cup \mathbf{H}, \mathbf{B}^*}(v) \cap \mathbf{H} \subseteq \mathbf{H}^*$.

By the DAG g-formula and the above features of the construction,
\begin{equation}
    \begin{split}
         p(\mathbf{Y}^* &| \textit{do}_{\mathcal{G}(\mathbf{O} \cup \mathbf{H}, \mathbf{B}^*)}(\mathbf{a}^*))\\
         &= \sum_{\mathbf{H}} \prod_{v \in (\mathbf{H} \cup \mathbf{Y}^*)} p(v | \pa_{\mathcal{G}(\mathbf{O} \cup \mathbf{H}, \mathbf{B}^*)}(v))\\
         &= \sum_{\mathbf{H}^*} \prod_{v \in (\mathbf{H}^* \cup \mathbf{Y}^*)} p(v | \pa_{\mathcal{G}(\mathbf{O} \cup \mathbf{H}, \mathbf{B}^*)}(v)) \cdot \sum_{\mathbf{H} \setminus \mathbf{H}^*} \prod_{v \in (\mathbf{H} \setminus \mathbf{H}^*)} p(v | \pa_{\mathcal{G}(\mathbf{O} \cup \mathbf{H}, \mathbf{B}^*)}(v))\\
         &= \sum_{\mathbf{H}^*} \prod_{\mathbf{D} \in \mathcal{D}(\mathcal{G}^*(\mathbf{Y}^*, \mathbf{A}^* \cup \mathbf{B}^*))} \prod_{v \in (\mathbf{D} \cup \mathbf{H}_{\mathbf{D}})} p(v | \pa_{\mathcal{G}(\mathbf{O} \cup \mathbf{H}, \mathbf{B}^*)}(v))\\
         &= \prod_{\mathbf{D} \in \mathcal{D}(\mathcal{G}^*(\mathbf{Y}^*, \mathbf{A}^* \cup \mathbf{B}^*))} \bigg( \sum_{\mathbf{H}_{\mathbf{D}}} \prod_{v \in (\mathbf{D} \cup \mathbf{H}_{\mathbf{D}})} p(v | \pa_{\mathcal{G}(\mathbf{O} \cup \mathbf{H}, \mathbf{B}^*)}(v)) \bigg).
    \end{split}
\end{equation}
For any district $\mathbf{D} \in \mathcal{D}(\mathcal{G}^*(\mathbf{Y}^*, \mathbf{B}^* \cup \mathbf{A}^*))$,
\begin{equation}
    \begin{split}
        \sum_{\mathbf{H}_{\mathbf{D}}} & \prod_{v \in \mathbf{D} \cup \mathbf{H}_{\mathbf{D}}} p(v | \pa_{\mathcal{G}(\mathbf{O} \cup \mathbf{H}, \mathbf{B}^*)}(v))\\
        &= \sum_{\mathbf{H}_{\mathbf{D}}} \prod_{v \in (\mathbf{D} \cup \mathbf{H}_{\mathbf{D}})} p(v | \pa_{\mathcal{G}(\mathbf{O} \cup \mathbf{H}, \mathbf{B}^*)}(v)) \cdot \sum_{\mathbf{H} \setminus \mathbf{H}_{\mathbf{D}}} \prod_{v \in (\mathbf{H} \setminus \mathbf{H}_{\mathbf{D}})} p(v | \pa_{\mathcal{G}(\mathbf{O} \cup \mathbf{H}, \mathbf{B}^*)}(v))\\
        &= \sum_{\mathbf{H}} \prod_{v \in \mathbf{D} \cup \mathbf{H}_{\mathbf{D}}} p(v | \pa_{\mathcal{G}(\mathbf{O} \cup \mathbf{H}, \mathbf{B}^*)}(v))\\
        &= \sum_{\mathbf{H}} \phi_{\mathbf{D}^* \setminus \mathbf{D}}(q(\mathbf{D}^* | \pa_{\mathcal{G}(\mathbf{O} \cup \mathbf{H}, \mathbf{B}^*)}(\mathbf{D}^*))); \mathcal{G}(\mathbf{O} \cup \mathbf{H}, \mathbf{B}^*))
    \end{split}
\end{equation}
Once again, these equalities are a result of the above constructions of $\mathbf{H}$ and $\mathbf{H}^*$. By commutativity (Lemma 55 in \cite{richardson17nested}), we can remove references to $\mathbf{H}$:
\begin{align*}
    p(\mathbf{Y}^* | &\textit{do}_{\mathcal{G}(\mathbf{O} \cup \mathbf{H}, \mathbf{B}^*)}(\mathbf{A}^*))\\
    &= \prod_{\mathbf{D} \in \mathcal{D}(\mathcal{G}(\mathbf{Y}^*, \mathbf{B}^* \cup \mathbf{A}^*))} \phi_{\mathbf{D}^* \setminus \mathbf{D}} q(\mathbf{D}^* | \pa_{\mathcal{G}(\mathbf{O}, \mathbf{B}^*)}(\mathbf{D}^*)); \mathcal{G}(\mathbf{O}, \mathbf{B}^*))\\
    &= \prod_{\mathbf{D} \in \mathcal{D}(\mathcal{G}(\mathbf{Y}^*, \mathbf{B}^* \cup \mathbf{A}^*))} \phi_{\mathbf{D}^* \setminus \mathbf{D}} q(\mathbf{D}^* | \pa_{\mathcal{G}}(\mathbf{D}^*)); \mathcal{G}^d)\\
    &= \prod_{\mathbf{D} \in \mathcal{D}(\mathcal{G}_{\mathbf{Y}^*})} \phi_{\mathbf{D}^* \setminus \mathbf{D}} q(\mathbf{D}^* | \pa_{\mathcal{G}}(\mathbf{D}^*)); \mathcal{G}^d)
\end{align*}
The second equality is true because $\pa_{\mathcal{G}}(\mathbf{D}^*) \subseteq \pa_{\mathcal{G}(\mathbf{O}, \mathbf{B}^*)}(\mathbf{D}^*)$ and by the assumption of a block-safe chain graph. The final equality is true by block-safeness and the definition of induced subgraphs.

Finally by the fact that $p(\mathbf{Y} | \textit{do}_{\mathcal{G}(\mathbf{O} \cup \mathbf{H}, \mathbf{B}^*)}(\mathbf{A})) = \sum_{\mathbf{Y}^* \setminus \mathbf{Y}} p(\mathbf{Y}^* | \textit{do}_{\mathcal{G}(\mathbf{O} \cup \mathbf{H}, \mathbf{B}^*)}(\mathbf{A}^*))$, we can re-write the above as:
\begin{align*}
    p(\mathbf{Y} | \textit{do}_{\mathcal{G}(\mathbf{O} \cup \mathbf{H}, \mathbf{B}^*)}(\mathbf{A})) = \sum_{\mathbf{Y}^* \setminus \mathbf{Y}} \prod_{\mathbf{D} \in \mathcal{D}(\mathcal{G}_{\mathbf{Y}^*})} \phi_{\mathbf{D}^* \setminus \mathbf{D}} q(\mathbf{D}^* | \pa_{\mathcal{G}}(\mathbf{D}^*)); \mathcal{G}^d)
\end{align*}
We combine this with the block portioned derived above via chain-graph g-formula to obtain the result of the sub-claim

\textbf{Claim 3}: \textit{If there is a district in $\mathcal{D}(\mathcal{G}(\mathbf{O})_{\mathbf{Y}^*})$ that is not reachable in $\mathcal{G}^d$, then $p(\mathbf{Y}|\textit{do}(\mathbf{a}))$ is not identifiable.}

Let $\mathbf{D} \in \mathcal{D}(\mathcal{G}(\mathbf{O})_{\mathbf{Y}^*})$ be unreachable. Let $\mathbf{R} = \{D \in \mathbf{D} | \ch_{\mathcal{G}}(D) \cap \mathbf{D} = \emptyset\}$. Let $\mathbf{A}^* = A \cap \pa_{\mathcal{G}}(D)$. Then there exists a superset of $\mathbf{D}$, $\mathbf{D}'$, such that $\mathbf{D}$ and $\mathbf{D}'$ form a hedge for $p(\mathbf{R}|\textit{do}(\mathbf{a}^*))$ and thus $p(\mathbf{R}|\textit{do}(\mathbf{a}^*))$ is not identified \cite{shpitser06id}.

Let $\mathbf{Y}'$ be the minimal subset of $\mathbf{Y}$ such that $\mathbf{R} \subseteq \ant_{\mathcal{G}(\mathbf{O})_{\mathbf{O} \setminus \mathbf{A}}}(\mathbf{Y}')$. Consider an edge subgraph $\mathcal{G}^{\dag}$ of $\mathcal{G}$ consisting of all edges in $\mathcal{G}$ in the hedge formed by $\mathbf{D}, \mathbf{D}'$ and edges on partially directed paths in $\mathcal{G}(\mathbf{O})_{\mathbf{O} \setminus \mathbf{A}}$ from every element in $\mathbf{R}$ to some element in $\mathbf{Y}'$, such that the edge subgraph does not contain any cycles (directed or otherwise).

We proceed as follows.  We first define an ADMG $\tilde{\cal G}^{\dag}$ from ${\cal G}^{\dag}$ as follows.  The vertices and edges making up the hedge structure \cite{shpitser06id} in ${\cal G}^{\dag}$ are also present in $\tilde{\cal G}^{\dag}$.  For every partially directed path $\sigma$ from an element in ${\bf R}$ to an element in ${\bf Y}'$, we construct a directed path from ${\bf R}$ in $\tilde{\cal G}^{\dag}$ containing vertex copies of vertices on the undirected path $\sigma$, and which orients all undirected edges in $\sigma$ away from ${\bf R}$ and towards the element copy in $\tilde{\cal G}^{\dag}$ of the appropriate element of ${\bf Y}'$ in ${\cal G}^{\dag}$.

We then prove non-identifiability of $p(\tilde{\bf Y}' | \text{do}({\bf a}^*))$ in $\tilde{\cal G}^{\dag}$, where $\tilde{\bf Y}'$ is the set of all vertex copies in $\tilde{\cal G}^{\dag}$ of vertices in ${\bf Y}'$ in ${\cal G}^{\dag}$, using standard techniques for ADMGs.  In particular, we follow the proof of Theorem 4 in the supplement of \cite{shpitser18medid}.

We next show that $p({\bf Y}' \mid \text{do}({\bf a}^*))$ is not identified in ${\cal G}^{\dag}$.  For the two counterexamples in the causal model given by $\tilde{\cal G}^{\dag}$ witnessing non-identifiability of $p(\tilde{\bf Y}' \mid \text{do}({\bf a}^*))$ in the above proof, we will construct two counterexamples in the causal model given by ${\cal G}^{\dag}$ witnessing non-identifiability of $p({\bf Y}' \mid \text{do}({\bf a}^*))$.

To do so, we define new variables along all partially directed paths from ${\bf R}$ to ${\bf Y}'$ in ${\cal G}^{\dag}$ as Cartesian products of variable copies in counterexamples constructed.  Note that any such variable containing only a single element in ${\bf R}$ in its anterior in
${\cal G}^{\dag}$ will only have a single copy, while a variable containing two elements in ${\bf R}$ in its anterior in ${\cal G}^{\dag}$ will contain two copies, and so on.   It's clear that the two resulting elements contain vertices in ${\cal G}^{\dag}$, agree on the observed data distribution, and disagree on $p({\bf Y}' \mid \text{do}({\bf a}^*))$.

What remains to show is that the distributions so constructed obey one of CG Markov properties associated with a CG ${\cal G}^{\dag}$.
Fix a (possibly trivial) block ${\bf B}$ in ${\cal G}^{\dag}$.  We must show for each $B \in {\bf B}$ that
$p(B \mid {\bf B} \setminus B, \pa_{{\cal G}^{\dag}}({\bf B})) = p(B \mid \nb_{{\cal G}^{\dag}}, \pa_{\cal G}(B))$.

For any $B \in {\bf B}$ in ${\cal G}^{\dag}$, there exists a set $B_1, \ldots, B_k$ of variables in $\tilde{\cal G}^{\dag}$ such that
$B$ is defined as $B_1 \times \ldots \times B_k$.  Moreover, any variable $A \in \nb_{{\cal G}^{\dag}}(B) \cup \pa_{{\cal G}^{\dag}}(B)$ corresponds to a Cartesian product $A_1 \times A_m$ of variables where $A_i$ is a child or a parent of some variables $B_j$.
The result then follows by d-separation in $\tilde{\cal G}^{\dag}$, and the fact that the part of $\tilde{\cal G}^{\dag}$ outside of the hedge structure does not contain any colliders by construction.
\end{prf}

\section{Derivations}

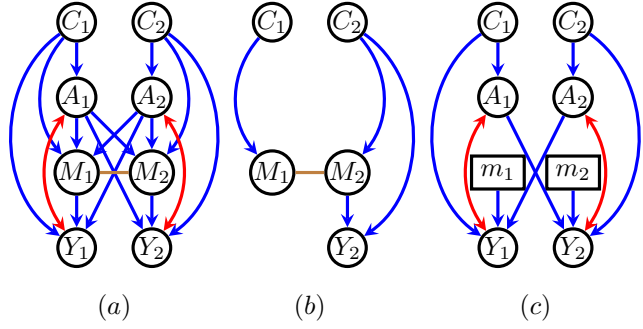
\begin{figure}[t]
	\begin{center}
		\begin{tikzpicture}[>=stealth, node distance=1.0cm]
		\tikzstyle{format} = [draw, very thick, circle, minimum size=5mm,
		inner sep=0pt]
		\tikzstyle{square} = [draw, very thick, rectangle, minimum size=3.8mm]

		\begin{scope}[xshift=5.0cm] 
		\path[->, very thick]
		node[format] (a1) {$A_1$}
		node[format, above of=a1] (c1) {$C_1$}
		node[format, below of=a1] (m1) {$M_1$}
		node[format, below of=m1] (y1) {$Y_1$}
		
		node[format, right of=a1] (a2) {$A_2$}
		node[format, above of=a2] (c2) {$C_2$}
		node[format, below of=a2] (m2) {$M_2$}
		node[format, below of=m2] (y2) {$Y_2$}
		
		
		(c1) edge[blue] (a1)
		(c1) edge[blue, bend right=40] (m1)
		(c1) edge[blue, bend right=55] (y1)
		(a1) edge[blue] (m1)
		(m1) edge[blue] (y1)
		(a1) edge[<->, red, bend right=35] (y1)
		
		(c2) edge[blue] (a2)
		(c2) edge[blue, bend left=40] (m2)
		(c2) edge[blue, bend left=55] (y2)
		(a2) edge[blue] (m2)
		(m2) edge[blue] (y2)
		(a2) edge[<->, red, bend left=35] (y2)
		
		(a1) edge[blue] (y2)
		(a2) edge[blue] (y1)
		
		
		(a1) edge[blue] (m2)
		(a2) edge[blue] (m1)
		
		(m1) edge[-, brown] (m2)


		node[below of=y1, yshift=0.2cm, xshift=0.5cm] (l) {$(a)$}
		;
		\end{scope}

		\begin{scope}[xshift=8.6cm] 
		\path[->, very thick]
		node[] (a2) {} 
		node[format, above of=a2] (c2) {$C_2$}
		node[format, below of=a2] (m2) {$M_2$}
		node[format, below of=m2] (y2) {$Y_2$}
		
		node[left of=a2] (a1) {}
		node[format, left of=c2] (c1) {$C_1$}
		node[format, below of=a1] (m1) {$M_1$}
		
		(c1) edge[blue, bend right=40] (m1)
		
		(c2) edge[blue, bend left=40] (m2)
		(c2) edge[blue, bend left=55] (y2)
		(m2) edge[blue] (y2)
		
		
		

		
		(m1) edge[-, brown] (m2)
		node[below of=y2, yshift=0.2cm, xshift=-0.5cm] (l) {$(b)$}
		;
		\end{scope}
		
		\begin{scope}[xshift=10.6cm]
		\path[->, very thick]
		node[format] (a1) {$A_1$}
		node[format, above of=a1] (c1) {$C_1$}
		node[square, below of=a1] (m1) {$m_1$}
		node[format, below of=m1] (y1) {$Y_1$}
		
		node[format, right of=a1] (a2) {$A_2$}
		node[format, above of=a2] (c2) {$C_2$}
		node[square, below of=a2] (m2) {$m_2$}
		node[format, below of=m2] (y2) {$Y_2$}
		
		(c1) edge[blue] (a1)
		(c1) edge[blue, bend right=55] (y1)
		(a1) edge[<->, red, bend right=35] (y1)
		
		(c2) edge[blue] (a2)
		(c2) edge[blue, bend left=55] (y2)
		(a2) edge[<->, red, bend left=35] (y2)
		
		(m1) edge[blue] (y1)
		(m2) edge[blue] (y2)
		
		(a1) edge[blue] (y2)
		(a2) edge[blue] (y1)


		node[below of=y1, yshift=0.2cm, xshift=0.5cm] (l) {$(c)$}
		;
		\end{scope}
		
		\end{tikzpicture}
	\end{center}
	\caption{
		(a) A latent projection of the CG in (Fig. 1a in the main paper) onto observed variables.
		(b) The graph representing $\mathcal{G}_{\mathbf{Y^*}}$ for the intervention operation $\text{do}(a_1)$ applied to (a).
		(c) The ADMG obtained by fixing $M_1, M_2$ in (a).
	}
	\label{fig:front_door}
\end{figure}

Consider Figure \ref{fig:front_door} (a). We are interested in identifying  
$p(Y_2(a_1,a_2))$. We set $\mathbf{Y^*}$ to the anterior of ${\bf Y}$ in $\mathcal{G}_{\mathbf{V} \setminus \mathbf{A}}$: $\mathbf{Y^*} \equiv \{C_1, C_2, M_1, M_2, Y_2\}$ (see $\mathcal{G}_{\mathbf{Y^*}}$ shown in Fig. \ref{fig:front_door} (b)) with $\mathcal{B}(\mathcal{G}_{\mathbf{Y}^*}) = \{\{M_1, M_2\}\}$ and $\mathcal{D}(\mathcal{G}_{\mathbf{Y}^*} = \{\{C_1\}, \{C_2\},  \{Y_2\}\}$. We can now proceed with the version of the ID algorithm for SGs. The CCG portion of the algorithm simply yields $p(M_1, M_2 | A_1 = a_1, A_2, C_1, C_2)$. Note that this expression further factorizes according to the factorization of blocks in a chain graph. For the ADMG portion of the algorithm, we must fix variables in three different sets $\{C_2, A_1, A_2, Y_1, Y_2\}$, $\{C_1, A_1, A_2, Y_1, Y_2\}$,
$\{C_1, C_2, A_1, A_2, Y_1\}$ in $\mathcal{G}^d$, shown in Fig. \ref{fig:front_door} (c), corresponding to three districts in Fig.~\ref{fig:front_door} (b).  We have:
\begin{equation}
    \begin{split}
        \phi_{\{C_2, A_1, A_2, Y_1, Y_2\}}(p(Y_1, Y_2 | A_1, A_2, M_1, M_2, C_1, C_2)p(A_1, A_2, C_1, C_2))\\
        = \phi_{\{C_2, A_1, A_2, Y_1\}}(p(Y_1 | A_1, A_2, M_1, M_2, C_1, C_2, Y_2)p(A_1, A_2, C_1, C_2))\\
        = \phi_{\{C_2, A_1, A_2\}}(p(A_1, A_2, C_1, C_2))\\
        = \phi_{\{C_2, A_2\}}(p(A_2, C_1, C_2))\\
        = \phi_{\{C_2\}}(p(C_1, C_2))\\
        = p(C_1)\\
    \end{split}
\end{equation}

\begin{equation}
    \begin{split}
        \phi_{\{C_1, A_1, A_2, Y_1, Y_2\}}(p(Y_1, Y_2 | A_1, A_2, M_1, M_2, C_1, C_2)p(A_1, A_2, C_1, C_2))\\
        = \phi_{\{C_1, A_1, A_2, Y_1\}}(p(Y_1 | A_1, A_2, M_1, M_2, C_1, C_1, Y_2)p(A_1, A_2, C_1, C_2))\\
        = \phi_{\{C_1, A_1, A_2\}}(p(A_1, A_2, C_1, C_2))\\
        = \phi_{\{C_1, A_2\}}(p(A_2, C_1, C_2))\\
        = \phi_{\{C_1\}}(p(C_1, C_2))\\
        = p(C_2)
    \end{split}
\end{equation}

\begin{equation}
    \begin{split}
        \phi_{\{C_1, C_2, A_1, A_2,Y_1\}}(p(Y_1, Y_2 | A_1, A_2, M_1, M_2, C_1, C_2)p(A_1, A_2, C_1, C_2))\\
        = \phi_{\{A_1, Y_1,A_2\}}(p(Y_1, Y_2 | A_1, A_2, M_1, M_2, C_1, C_2)p(A_1, A_2 | C_1, C_2))\\
        = \phi_{\{A_1,A_2\}}(p(Y_2 | A_1, A_2, M_1, M_2, C_1, C_2)p(A_1, A_2 | C_1, C_2))\\
        = \sum_{A_2} p(Y_2 | A_1, A_2, M_1, M_2, C_1, C_2)p(A_2 | C_2)\\
        = \sum_{A_2} p(Y_2 | A_1, A_2, M_2, C_2)p(A_2 | C_2)
    \end{split}
\end{equation}
with the last term evaluated at $A_1 = a_1$.  Thus, the identifying functional is:
\begin{equation}
\begin{split}
     p(Y_2(a_1,a_2)) = \sum_{\{C_1, C_2, M_1, M_2\}} &\bigg[p(M_1, M_2 | a_1, a_2, C_1, C_2) \\ 
    \times &\Big[\sum_{A_2} p(Y_2 | a_1,A_2, M_2, C_2)p(A_2|C_2) p(C_1) p(C_2)\Big] \bigg]
\end{split}
\label{eqn:result}
\end{equation}

\section{Simulation Study}

%

\subsection{The Auto-G-Computation Algorithm}

To estimate identifying functionals corresponding to causal effects given dependent data, we generally use maximum likelihood plug in estimation.  The exception is the factor $p({\bf M} \mid \pa_{\cal G}({\bf M}))$, which may not be estimated if $M_i$ variables for all units $i$ are dependent, as is the case in our simulation study.  In this case, the above density must be estimated from a single sample.  Thus, standard statistical methods such as maximum likelihood estimation fail to work.  We adapt the auto-g-computation algorithm method in
\cite{tchetgen2017auto}, which exploits Markov assumptions embedded in our CG model, as well as the pseudo-likelihood or coding estimation methods introduced in \cite{besag75pseudo}.  We briefly describe the approach here.

The auto-g-computation algorithm is a generalization of the Monte Carlo sampling version of the standard g-computation algorithm for classical causal models (represented by DAGs) \cite{westreich12parametric} to causal models represented by CGs.
Auto-g-computation proceeds by generating samples from a block using Gibbs sampling.  The parameters for Gibbs factors used in the sampler (which, by the global Markov property for CGs, take the form of
$p( X_i \mid \pa_{\cal G}(X_i) \cup \nb_{\cal G}(X_i))$) are learned via parameter sharing and coding or pseudo-likelihood based estimators.
For any block ${\bf B}$, the Gibbs sampler draws samples from $p(\mathbf{X} \mid \pa_{\cal G}({\bf X}))$, given a fixed set of samples drawn from all blocks with elements in $\pa_{\cal G}({\bf X})$, or specific values of $\pa_{\cal G}({\bf X})$ we are interested in, as follows.\\
\underline{Gibbs Sampler for ${\bf X}$}:
\begin{align*}
\text{for }t  &  =0,\text{let } \mathbf{x}^{(0)}  \text{ denote initial values ;}\\
\text{for }t  &  =1,...,T\\
&  \text{draw value of }X_{1}^{(t)}\text{ from }p(  X_{1} | {\bf x}^{(t-1)}_{\pa_{\cal G}(X_1) \cup \nb_{\cal G}(X_1)} )
);\\
&  \text{draw value of }X_{2}^{(t)}\text{ from }p(  X_{2} | {\bf x}^{(t-1)}_{\pa_{\cal G}(X_2) \cup \nb_{\cal G}(X_2)} )
);\\
&  \vdots\\
&  \text{draw value of }X_{m}^{(t)}\text{ from }p(  X_{m} | {\bf x}^{(t-1)}_{\pa_{\cal G}(X_m) \cup \nb_{\cal G}(X_m)} )
);
\end{align*}
Since we are interested in estimating a functional similar to (\ref{eqn:result}), we use observed values of ${\bf C}$, and intervened on values $a_i,a_j$ as the values of $\pa_{\cal G}({\bf M})$ in the Gibbs sampler.

The coding-likelihood and pseudo-likelihood estimators we use are described in more detail in \cite{tchetgen2017auto}.
Both estimators rely on parameter sharing for densities $p(M_i \mid \pa_{\cal G}(M_i) \cup \nb_{\cal G}(M_i))$ across different units $i$, and for the network to be sufficiently sparse such that each $M_i$ depends on only a few other variables in the model, relative to the total number of units.

The coding estimator uses a subset of the data that corresponds to units that form independent sets in the network adjacency graph (where units are adjacent of they are friends in the network, and not adjacent otherwise).
A set of units is a \emph{maximal} independent set in the network adjacency graph if a) no two vertices in the set are adjacent, and b) it is impossible to add another unit to the set without violating the adjacency constraint. A \emph{maximum} independent set is a maximal independent set such that there does not exist a larger maximal independent set in the same graph. Finding maximum independent sets is  a classic NP-complete problem; in practice we find several \emph{maximal} independent sets and pick the one with largest cardinality as a heuristic. See Table \ref{tab:s_max} below for the size of $S_{max}$ for each network size in our experiments.
\begin{table}[h]
\begin{center}
\begin{center}
\begin{tabular}{ |c|c|c|c|c| }
\hline
 $N$ & 400 & 800 & 1000 & 2000 \\ 
 \hline
 $|S_{max}|$ & 159 & 309 & 384 & 763 \\
 \hline
\end{tabular}
\end{center}
\end{center}
\caption{The size of $S_{max}$ used for the coding-likelihood estimator in each network}
\label{tab:s_max}
\end{table}
The coding likelihood estimator was proven consistent and asymptotically normal in \cite{tchetgen2017auto} whereas pseudo-likelihood estimation is, under mild assumptions, consistent but not asymptotically normal. On the other hand, pseudo-likelihood estimation is more efficient than coding likelihood estimation since it makes use of all of the data.

\subsection{Simulation Specifics}
For data generation we use the following densities for $A_i, M_i, Y_i$, parameterized by
$\tau_{A} = \{\gamma_{0}, \gamma_{C_{1}}, \dots, \gamma_{C_{p}}, \gamma_{U_{1}}, \dots, \gamma_{U_{q}}\}, \tau_{M} = \{\beta_{0}, \beta_{A}, \beta_{C_{1}}, \dots, \beta_{C_{p}} \beta_{A_{nb}}, \beta_{M_{nb}}\}, \tau_{Y} = \{\alpha_{0}, \alpha_{C_{1}}, \dots, \alpha_{C_{p}}, \alpha_{U_{1}}, \dots, \alpha_{U_{q}}, \alpha_{A_{nb}}, \alpha_{M}\}$:
\begin{align*}
    p(A_i = 1 &| \mathbf{C}_i, \mathbf{U}_i; \tau_{A}) = expit(\gamma_{0} + \big(\sum_{l = 1}^{p} \gamma_{C_{l}} C_{il}\big) + \big(\sum_{l = 1}^{q} \gamma_{U_{l}} U_{il}\big))\\
    p(M_i = 1 &| A_i, \mathbf{C}_i, \{A_j, M_j | j \in \mathcal{N}_{i}\}; \tau_{M})\\
    &= expit(\beta_{0} + \beta_{A} A_i + \big(\sum_{l = 1}^{p} \beta_{C_{l}} C_{il}\big) + \big(\sum_{j \in \mathcal{N}_{i}} (\beta_{A_{nb}} A_{j} + \beta_{M_{nb}} M_j)\big))\\
    p(Y_i = 1 &| \mathbf{C}_i, \mathbf{U}_i, M_i, \{A_j | j \in \mathcal{N}_j\}; \tau_{Y}) \\
    &= expit(\alpha_{0} 
    + \big(\sum_{l = 1}^{p} \alpha_{C_{l}} C_{il}\big) + \big(\sum_{l = 1}^{q} \alpha_{U_{l}} U_{il}\big)
    + \big(\sum_{j = \mathcal{N}_i} \alpha_{A_{nb}} A_{j}\big) + \alpha_{M}M_i).
\end{align*}

The values of the parameters for the beta distributions we use to generate $\mathbf{C}_i, \mathbf{U}_i$ can be found in Table \ref{tab:CUparams} while the values of $\tau_A, \tau_M, \tau_Y$ can be found in Table \ref{tab:tauparams}.

\begin{table}[h]
    \centering
    \subfloat[Parameters for $\mathbf{C}$ and $\mathbf{U}$]{
    \begin{tabular}{|c|c|c|}
        \hline
        Variable & a & b \\
        \hline
        $C_1$ & 1.5 & 3 \\
        \hline
        $C_2$ & 6 & 2 \\
        \hline
        $C_3$ & 0.8 & 0.8 \\
        \hline
        $U_1$ & 2.3 & 1.1 \\
        \hline
        $U_2$ & 0.9 & 1.1 \\
        \hline
        $U_3$ & 2 & 2 \\
        \hline
    \end{tabular}\label{tab:CUparams}}
    \hspace{.5cm}
    \subfloat[Parameters for $\tau_{A}, \tau_{M}, \tau{Y}$]{
    \begin{tabular}{|c|c|}
        \hline
        Parameter & Value \\
        \hline
        $\tau_{A}$ & (-1, 0.5, 0.2, 0.25, 0.3, -0.2, 0.25)\\
        \hline
        $\tau_{M}$ & (-1, -0.3, 0.4, 0.1, 1, -0.5, -1.5) \\
        \hline
        $\tau_{Y}$ & (-0.3, -0.2, 0.2, -0.05, 0.1, -0.2, 0.25, -1, 3) \\
        \hline
    \end{tabular}\label{tab:tauparams}}
    \caption{The parameters for each generating distribution}
\end{table}

\subsection{Extended Results}
In the main paper we gave confidence intervals and the mean and standard deviation of the bias of our estimators. All results were calculated by averaging over $1000$ simulated networks.
\begin{table}[h]
\begin{center}
\begin{tabular}{|*5{m{15mm}|}}
\hline
\multicolumn{5}{|c|}{Ground Truth Network Average Effects}\\
\hline
$N$ & 400 & 800 & 1000 & 2000\\
\hline
Ground Truth &-.455& -.453& -.455& -.456\\
\hline
\end{tabular}
\end{center}
\caption{The ground truth effects for each network, calculated by averaging over 5 samples of the data generating process for each network under the relevant interventions}
\label{tab:ground_truth}
\end{table}

As discussed in the main body of the paper, the estimators we use are able to recover the effects of interest reasonably well. The approximate ground truth values for these effects can be found in Table \ref{tab:ground_truth}. The fact that the coding estimator restricts the network to a small fraction of its total units means it is considerably less efficient than the pseudo-likelihood estimator.

Though the pseudo-likelihood estimator is not in general asymptotically normal, it does not perform substantially worse than the provably asymptotically normal coding-likelihood estimator. In both cases, the true effect is covered by the 95\% confidence interval of the estimator. 
%

\end{document}